\documentclass[a4]{article}
\usepackage[english]{babel}
\usepackage{authblk}
\usepackage{booktabs} 
\usepackage{arydshln}
\usepackage{amsmath, amssymb, amsfonts}
\usepackage{tikz}
\usepackage{pgfplots}
\pgfplotsset{compat=1.11}
\usepackage{graphicx}
\usetikzlibrary{arrows.meta, svg.path}
\usepackage[linesnumbered, ruled, longend]{algorithm2e}
\usepackage[config, position=bottom]{subfig}
\usepackage[pdftex,
            pdfauthor={Dickson Owuor, Thomas Runkler, Anne Laurent, Edmond Menya},
            pdftitle={Ant Colony Optimization of Mining Gradual Patterns},
            pdfsubject={Gradual Patterns},
            pdfkeywords={pheromone}]{hyperref}
\usepackage{multirow}
\usepackage{rotating}

	\usepackage{filecontents}

\author[1]{Dickson Odhiambo Owuor}
\author[2]{Thomas Runkler}
\author[3]{Anne Laurent}
\author[1]{Joseph Onderi Orero}
\author[1]{Edmond Odhiambo Menya}

\affil[1]{SCES Strathmore University, Nairobi, Kenya\\ 

	\textit{\{dowuor, jorero, emenya\}@strathmore.edu}}
\affil[2]{Siemens AG, Munich, Germany\\ 

	\textit{thomas.runkler@siemens.com}}
\affil[3]{LIRMM Univ Montpellier, CNRS, Montpellier, France\\

	\textit{anne.laurent@umontpellier.fr}}

\title{Ant Colony Optimization for Mining Gradual Patterns}

\begin{document}
\maketitle

\begin{abstract}
Gradual pattern extraction is a field in Knowledge Discovery in Databa-ses that maps correlations between attributes of a data set as gradual dependencies. A gradual dependency may take the form: \textit{``the more Attribute$_{K}$, the less Attribute$_{L}$''}. Classical approaches for extracting gradual patterns extend either a breath-first search or a depth-first search strategy. However, these strategies can be computationally expensive and inefficient especially when dealing with large data sets. In this study, we investigate 3 population-based optimization techniques (i.e. ant colony optimization, genetic algorithm and particle swarm optimization) that may be employed improve the efficiency of mining gradual patterns. We show that ant colony optimization technique is better suited for gradual pattern mining task than the other 2 techniques. Through computational experiments on real-world data sets, we compared the computational performance of the proposed algorithms that implement the 3 population-based optimization techniques to classical algorithms for the task of gradual pattern mining and we show that the proposed algorithms outperform their classical counterparts.
\end{abstract}


	\section{Introduction}
	\label{sec1:intro}
	Knowledge Discovery in Databases (KDD) is a Computer Science field that attempts to unearth hidden information from huge data sets, which may otherwise take human analysts so long to discover. Gradual pattern (GP) discovery is a descriptive extension in the KDD field that expresses attribute correlations linguistically in the form of gradual rules \cite{Aryadinata2013,Aryadinata2014a,Berzal2007,Di-Jorio2009,Laurent2009}.
	
	Mining gradual dependencies may lead to uncovering new knowledge about numeric data sets in various kinds of domains. For instance, in a medical domain useful knowledge may be extracted through gradual rules like: \textit{``the faster the population growth of bacteria X, the slower the population growth of bacteria Y in organ G''} - may be denoted as $\{(bacteriaX, \uparrow), (bacteriaY, \downarrow)\}$.

	In order to formulate such causality, a candidate rule is generated from respective attribute combinations and its quality validated through a frequency support measure (which is derived from the sequence of all ordered tuples respecting the rule). For example the data set in Table~\ref{tab1:sample_set} has 3 attributes \{Game, Win, Injury\} from which we can formulate gradual rules like $\{(Game,\uparrow), (Win, \downarrow)\}$ (which is respected by at least 3/5 tuples), $\{(Game, \downarrow), (Win, \downarrow), (Injury, \uparrow)\}$ (which is respected by at least 2/5 tuples) etc.
		
	\begin{table}[h!]
  		\centering
    	\small
    	\caption{A sample data set containing number of: games, wins and injuries of an athlete.}
    	\begin{tabular}{c c c c} 
      	\textbf{id} & \textbf{Game} & \textbf{Win} & \textbf{Injury}\\
      	\hline \hline
      	r0 & 30 & 3 & 1\\
      	r1 & 35 & 2 & 2\\
      	r2 & 40 & 4 & 2\\
      	r3 & 50 & 1 & 1\\
      	r4 & 52 & 7 & 1\\
      	\bottomrule
    	\end{tabular}
    	\label{tab1:sample_set}
	\end{table}
	
	Many GP mining approaches extend either a \textit{breadth-first search} (BFS) or \textit{depth-first search} (DFS) strategy for mining gradual item sets. In this study, we propose an \textit{ant colony optimization} (ACO) which uses a probabilistic approach to improve efficiency of both BFS-based and DFS-based strategies for gradual item set mining. ACO, as originally described by \cite{Dorigo1996}, is a general-purpose heuristic approach for solving discrete optimization problems \cite{Blum2005,Smith2001,Dorigo2019,Runkler2005,Silva2002}.

	ACO imitates the positive feedback reinforcement behavior of biological ants as they search for food - where the more ants following a path, the more chemical pheromones are deposited on that path and the more appealing that path becomes for being followed by other ants \cite{Dorigo2010,Dorigo1996}.
	
\begin{figure}[h!]
		\centering
		\small
  		\subfloat[]{%
		\begin{tikzpicture}
		\begin{scope}
			\node (n1) at (0,1) {A};
			\node (n2) at (0,0) {B};
			\node (n3) at (0.8,-1) {C};
			\node (n4) at (-1.5,-1) {D};
			\node (n5) at (0,-2) {E};
			\node (n6) at (0,-3) {F};
		\end{scope}
	
		\begin{scope}
			\path (n1) edge[draw=black] (n2);
			\draw[-] (n2) edge  node[sloped, above, brown] {\scriptsize d=1} (n3);
			\draw[-] (n2) edge  node[sloped, above, brown] {\scriptsize d=2} (n4);
			\draw[-] (n3) edge  node[sloped, below, brown] {\scriptsize d=1} (n5);
			\draw[-] (n4) edge  node[sloped, below, brown] {\scriptsize d=2} (n5);
			\path (n5) edge[draw=black] (n6);
		\end{scope}
		\end{tikzpicture}
		}%
		$\qquad \qquad$
  		\subfloat[]{%
		\begin{tikzpicture}
		
		\begin{scope}[every node/.style={thin, draw}]
			\node (t) at (-1.2, 1.5) {t=0};
		\end{scope}
		
		\begin{scope}
			\node (n1) at (0,1) {A};
			\node (n2) at (0,0) {B};
			\node (n3) at (0.8,-1) {C};
			\node (n4) at (-1.5,-1) {D};
			\node (n5) at (0,-2) {E};
			\node (n6) at (0,-3) {F};
		\end{scope}
	
		\begin{scope}
			\path (n2) edge[draw=black] (n3);
			\path (n2) edge[draw=black] (n4);
			\path (n3) edge[draw=black] (n5);
			\path (n4) edge[draw=black] (n5);
			\path (n5) edge[draw=black] (n6);

			\draw[-] (n2) -- node[sloped, above, blue] {\scriptsize $\xleftarrow{12 ants}$}  (n1);
			\draw[dashed, blue] ([xshift=2.5ex]n2.south) edge[->]  node[sloped, above, blue] {\scriptsize 6 ants} (n3.north);
			\draw[dashed, blue] ([xshift=-4ex]n2.south) edge[->]  node[sloped, above, blue] {\scriptsize 6 ants} ([xshift=1ex]n4.north);
			
			\draw[-] (n5) -- node[sloped, above, red] {\scriptsize $\xleftarrow{12 ants}$}  (n6);
			\draw[dashed, red] ([xshift=2.5ex]n5.north) edge[->]  node[sloped, below, red] {\scriptsize 6 ants} ([yshift=0.2ex]n3.south);
			\draw[dashed, red] ([xshift=-4ex]n5.north) edge[->]  node[sloped, below, red] {\scriptsize 6 ants} ([xshift=1ex]n4.south);
		\end{scope}
		
		\end{tikzpicture}
		}%
		$\qquad \qquad$
  		\subfloat[]{%
		\begin{tikzpicture}
		
		\begin{scope}[every node/.style={thin, draw}]
			\node (t) at (-1.2, 1.5) {t=1};
		\end{scope}
		
		\begin{scope}
			\node (n1) at (0,1) {A};
			\node (n2) at (0,0) {B};
			\node (n3) at (0.8,-1) {C};
			\node (n4) at (-1.5,-1) {D};
			\node (n5) at (0,-2) {E};
			\node (n6) at (0,-3) {F};
		\end{scope}
	
		\begin{scope}
			\path (n2) edge[draw=black, very thick] (n3);
			\path (n2) edge[draw=gray, very thin] (n4);
			\path (n3) edge[draw=black, very thick] (n5);
			\path (n4) edge[draw=gray, very thin] (n5);
			\path (n5) edge[draw=black] (n6);

			\draw[-] (n2) -- node[sloped, above, blue] {\scriptsize $\xleftarrow{12 ants}$}  (n1);
			\draw[dashed, blue] ([xshift=2.5ex]n2.south) edge[->]  node[sloped, above, blue] {\scriptsize 8 ants} (n3.north);
			\draw[dashed, blue] ([xshift=-4ex]n2.south) edge[->]  node[sloped, above, blue] {\scriptsize 4 ants} ([xshift=1ex]n4.north);
			
			\draw[-] (n5) -- node[sloped, above, red] {\scriptsize $\xleftarrow{12 ants}$}  (n6);
			\draw[dashed, red] ([xshift=2.5ex]n5.north) edge[->]  node[sloped, below, red] {\scriptsize 8 ants} ([yshift=0.2ex]n3.south);
			\draw[dashed, red] ([xshift=-4ex]n5.north) edge[->]  node[sloped, below, red] {\scriptsize 4 ants} ([xshift=1ex]n4.south);
		\end{scope}
		
		\end{tikzpicture}
		}%
    	\caption{An example of artificial ants: (a) initial paths with weighted distances, (b) at time $t=0$ there is no pheromone intensity on any path, (c) at time $t=1$ the pheromone intensity is stronger on the shorter paths.}
  	\label{fig1:artificial_ants}
  	\end{figure}
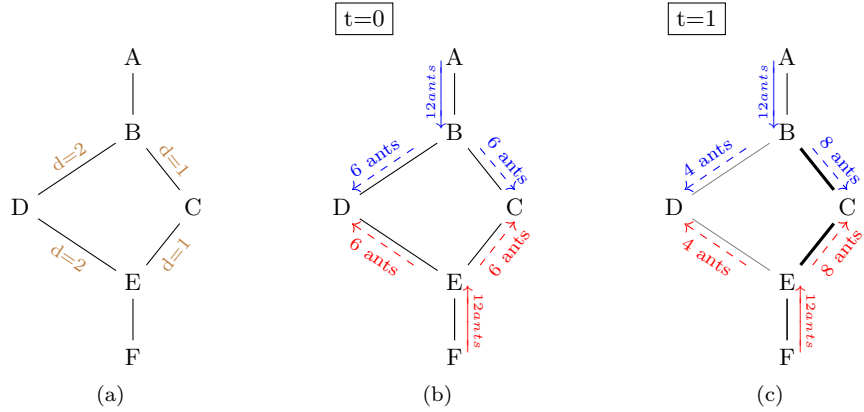

	\textit{Example 1.1.} We consider a sample graph of artificial ants moving along on the edges of nodes A, B, C, D, E and F as shown in Figure~\ref{fig1:artificial_ants}.
	
	In order to make an accurate interpretation of ACO, assume that the paths between nodes B and D, D and E have longer lengths than paths between nodes B and C, C and E (as indicated by the weighted distances in Figure~\ref{fig1:artificial_ants}a). Let us consider what happens at regular discrete time intervals: $t=0,1,2...$ . Suppose that 12 new ants come to node B from A and 12 new ants come to node E from F at each time interval. Each ant travels at a speed of \textit{2 weighted distance per time interval} and that by moving along at time $t$ it deposits pheromones of intensity 1, which completely and instantaneously evaporates in the middle of the successive time interval $(t+1, t+2)$.

	At time $t=0$ no path has any pheromone intensity and there are have 12 new ants at node B and 12 new ants at node E. The ants select between nodes C and D randomly; therefore, on average 6 ants from each node move towards C and 6 ants towards D (Figure~\ref{fig1:artificial_ants}b).
	
	At time $t=1$ there are 12 new ants at node B and 12 new ants at node E. The 12 new ants at node B will find a pheromone intensity of 6 on the path that leads to node D deposited by the 6 ants that used it previously from node B and a pheromone intensity of 12 on the path that leads to node C deposited by 6 ants that used it coming from node B and 6 that arrived from node E. Therefore, the ants will find the path to node C more desirable than that to node D. The probability of the ants choosing to move towards node C is 2/3, while that of moving towards node D is 1/3 and 8 ants will move towards node C and 4 ants towards node D. The same is true for the 12 new ants at node E (Figure~\ref{fig1:artificial_ants}c).
	
	\subsection*{Main Contributions}
	\label{sec1.1:contributions}
	The main contributions of the study are listed below:
	
	\begin{itemize}
		\item An ACO approach to BFS problem in GP mining.
		\item An ACO approach to DFS problem in GP mining.
		\item Alternative population-based optimization techniq-ues for GP mining.
	\end{itemize}
	
	Our experiment results show that algorithms built on top of population-based optimization techniques perform computationally better than their classical counterparts.
	
	The remainder of the paper is organized as follows: we provide preliminary concepts in Section~\ref{sec2:preliminary}; we review related literature in Section~\ref{sec3:related}; we propose ant colony optimization for GP extraction in Section~\ref{sec4:aco}; we show the results of our experiments in Section~\ref{sec5:experiments}. Finally, we conclude and give future research directions in Section~\ref{sec6:conclusion}.

	\section{Preliminary Concepts and Notations}
	\label{sec2:preliminary}	
	
	\subsection{Gradual Patterns}
	\label{sec2.2:gps}
		In this section, we provide some preliminary definitions of gradual item sets (as given by \cite{Di-Jorio2009,Laurent2009,Owuor2019}). In order to describe gradual patterns, we assume a data set $\mathcal{D}_{g}$ is defined by attributes $\{A_{1}, A_{2}, ..., A_{n}\}$ and it consists of tuples $\{r_{0}, r_{2}, ..., r_{n}\}$ as shown in Table~\ref{tab1:sample_set}.
	
	\textbf{Definition 2.1} \texttt{(Gradual Item)}. \textit{A gradual item is a pair $(i,v)$ where $i$ is an attribute and $v$ is a variation $v \in \lbrace \uparrow,\downarrow \rbrace$. $\uparrow$ stands for an increasing variation while $\downarrow$ stands for a decreasing variation}. 
	
	For example, $(wins,\uparrow)$ can be interpreted as ``the more wins''.
	
	\textbf{Definition 2.2} \texttt{(Gradual Pattern)}. \textit{A gradual pa-ttern $GP$ is a set of gradual items}:\\ $GP = \lbrace (i_{1},v_{1}), ..., (i_{n},v_{n})\rbrace$.
	
	For example, $\lbrace (games,\uparrow),(wins,\uparrow),(losses,\downarrow) \rbrace$ is a GP that can be interpreted as \textit{``the more games, the more wins, the less losses''}.
	
	In the following, let a data set $\mathcal{D}$ with attribute $A$ and $n$ tuples and $x$ be a tuple such that: $x \in \mathcal{D}$ and $A(x)$ denotes the value $A$ takes for $x$. A gradual dependency $(A, \uparrow)$ \textit{(the more $A$)} holds if $\forall x,x' \in \mathcal{D}, ~ A(x) < A(x')$ and vice-versa for $(A, \downarrow)$. The quality of gradual item sets is measured by \textit{frequency support}. The frequency support $sup$ of a GP is: \textit{``the proportion of tuple couples that respect the constraints by all the gradual items in the pattern''}, as given by the formula:
	
	\begin{footnotesize}
	\begin{equation}\label{eqn2:support}
		sup(GP) = \frac{1}{\mid \mathcal{D}' \mid}\mid \lbrace (x,x') \in \mathcal{D}' / \forall j \in [1,n] A_{j}(x) *_{j} A_{j}(x') \rbrace \mid
	\end{equation}
	\end{footnotesize}
	where $GP = \lbrace (A_{1},v_{1}), ..., (A_{n},v_{n})\rbrace$, $* \in \{<, >\}$ and $\mathcal{D}'$ is transaction data set (derived from $\mathcal{D}$) defined by attributes' gradual dependencies $(A_{n},v_{n})$ with the corresponding tuple pairs $(x,x')$ that respect them.
	
	Similarly, support describes the extent to which a GP holds for a given data set. Therefore, given a user-specified threshold $\sigma$ a GP is said to be \textit{frequent} if:
	
	\begin{equation}\label{eqn2:frequent}
		sup(GP) \geq \sigma
	\end{equation}
	
	\clearpage
	
	\subsection{Gradual Rule Traversal Techniques}
	\label{sec2.3:gradual_traversal}
	
	With regards to GP mining, two aspects complicate its mining process: (1) determining frequency support and (2) the \textit{complementary notion} of gradual item sets (for each attribute, there exist two gradual item sets). These two challenging aspects are brought about by the nature of patterns and thus traversal because we need to compare lines. As can be seen in Table~\ref{tab2:sample1} a and b, association rule mining deals with the transactions of a data set while gradual rule mining deals with the attributes of a data set.
	
	\begin{table}[h!]
  		\centering
  		\small
  		\caption{(a) Sample transactional data set (b) sample numeric data set.}
		\subfloat[]{%
    		\begin{tabular}{l l} 
      		\textbf{id} & \textbf{items}\\
      		\hline \hline
      		t1 & $\{d, a, c, b\}$\\
      		t2 & $\{a, d\}$\\
      		t3 & $\{b, c, a\}$\\
      		t4 & $\{d, a, c\}$\\
      		\bottomrule
    		\end{tabular}}%
    	\qquad \qquad
		\subfloat[]{%
    		\begin{tabular}{l c c c c}
      		\textbf{id} & \textbf{a} & \textbf{b} & \textbf{c} & \textbf{d} \\
      		\hline \hline
      		r1 & 5 & 30 & 43 & 97\\
      		r2 & 4 & 35 & 33 & 86\\
      		r3 & 3 & 40 & 42 & 108\\
      		r4 & 1 & 50 & 49 & 27\\
      		\bottomrule
    		\end{tabular}}%
    		\label{tab2:sample1}
	\end{table}
	
	To elaborate, in association rule mining a single transaction is enough to determine the occurrence frequency of an item set. For instance in Table~\ref{tab2:sample1}a given transaction \textit{t3} only, we can tell that item set \texttt{d} is not frequent. In gradual rule mining at least two or more transactions are needed to determine the frequency occurrence of an item set. For instance in Table~\ref{tab2:sample1}b given transaction \textit{r3} only, either $(a, \uparrow)$ or $(a, \downarrow)$ is possible. Further in order to mine gradual item sets, the \textit{complementary notion} requires that: for each attribute, there exist two gradual item sets. For instance attribute $a$ creates gradual item sets $(a, \uparrow)$ and $(a, \downarrow)$.
	
	In spite of this, techniques have been developed that allow for gradual rule mining through BFS and DFS strategies. In the case of BFS strategy for gradual rule mining, \cite{Di-Jorio2009,Laurent2009} propose approaches GRITE (GRadual ITemset Extraction) and GRAANK (GRAdual rANKing) that represent a gradual item set as a \textit{binary matrix of orders} and this enables the use of a level-wise BFS technique to generate \textit{$k$-itemset} gradual candidates.
	
	For example using Table~\ref{tab2:sample1}b, let us consider a $1$\textit{-item-set} gradual rule $i1 = (a,\downarrow )$. We have the list of tuples respecting this item set as $G_{i1} = \{(r1,r2), (r1,$\\$r3), (r1,r4), (r2,r3), (r2,r4), (r3,r4) \}$. The set of orders may be modeled using a binary matrix of size $4\times 4$ (where tuple pairs respecting the rule take the value of $1$ and those not respecting take the value of $0$) as illustrated in Table~\ref{tab2:binary_matrices}a. Table~\ref{tab2:binary_matrices}b shows the binary matrix of gradual item $(b,\uparrow )$.
	
	\begin{table}[h!]
		\centering
		\small
		\caption{Binary matrices $M_{G_{i1}}$, $M_{G_{i2}}$ for gradual items: (a) $i1 = (a,\downarrow )$, (b) $i2 = (b,\uparrow )$.}
		\subfloat[]{%
			\begin{tabular}{|c|c c c c|}
			\hline
			$\Rsh$ & r1 & r2 & r3 & r4\\
			\hline
			r1 & 0 & 1 & 1 & 1\\
			r2 & 0 & 0 & 1 & 1\\
			r3 & 0 & 0 & 0 & 1\\
			r4 & 0 & 0 & 0 & 0\\
			\hline
			\end{tabular}
			$~~~$ 
			}%
		\subfloat[]{%
			\begin{tabular}{|c|c c c c|}
			\hline
			$\Rsh$ & r1 & r2 & r3 & r4\\
			\hline
			r1 & 0 & 1 & 1 & 1\\
			r2 & 0 & 0 & 1 & 1\\
			r3 & 0 & 0 & 0 & 1\\
			r4 & 0 & 0 & 0 & 0\\
			\hline
			\end{tabular}
			$~~~$ 
			}%
		\label{tab2:binary_matrices}
	\end{table}
	
	In the case of DFS strategy for gradual rule mining, \cite{Negrevergne2014} extends LCM (Linear time Closed item set Miner) presented by \cite{Uno2003,Uno2004} to propose ParaMiner. Before applying a DFS technique to build a (FP-Tree) \textit{frequent pattern tree} (a set enumeration tree of long frequent items recursively grown from short ones) of gradual item sets; ParaMiner encodes a numeric data set into transactional data set containing gradual dependencies of attributes with the corresponding concordant pairs as shown in Table~\ref{tab2:gp_encoding}a.
	
	Both GRAANK, and ParaMiner techniques have their merits and demerits. GRAANK's relies on the bitmap representation of gradual variations using a binary matrix. This makes arithmetic computations of candidates very efficient. ParaMiner employs a parallel algorithm which relies on a data set reduction technique to improve its efficiency. Table~\ref{tab2:gp_encoding}b illustrates how a data set is reduced by grouping similar gradual items (with weights) and removing those whose support do not surpass the user-specified threshold.
	
	\begin{table}[h!]
		\centering
		\small
		\caption{(a) Transactional encoding of data set in Table~\ref{tab2:sample1}b, (b) reduced data set of the encoded data set (where \texttt{tids} - \textbf{t}ransaction \textbf{id}s with similar item sets  and \texttt{weight} - number of appearance).}
		\subfloat[]{%
		\begin{tabular}{l l }
      		\textbf{id} & \textbf{item-sets} \\
      		\hline \hline
      		$t_{(r1, r2)}$ & $\{(a,\downarrow), (b,\uparrow), (c,\downarrow), (d,\downarrow) \}$\\
      		$t_{(r1, r3)}$ & $\{(a,\downarrow), (b,\uparrow), (c,\downarrow), (d,\uparrow) \}$\\
      		$t_{(r1, r4)}$ & $\{(a,\downarrow), (b,\uparrow), (c,\uparrow), (d,\downarrow) \}$\\
      		$t_{(r2, r3)}$ & $\{(a,\downarrow), (b,\uparrow), (c,\uparrow), (d,\uparrow) \}$\\
      		$t_{(r2, r4)}$ & $\{(a,\downarrow), (b,\uparrow), (c,\uparrow), (d,\downarrow) \}$\\
      		$t_{(r3, r4)}$ & $\{(a,\downarrow), (b,\uparrow), (c,\uparrow), (d,\downarrow) \}$\\
      		\bottomrule
    	\end{tabular}
    	}%
    	\\
		\subfloat[]{%
		\begin{tabular}{l c l}
      		\textbf{tids} & \textbf{weight} & \textbf{item-sets} \\
      		\hline \hline
      		$t_{(r1, r2)}$ & 1 & $\{(a,\downarrow), (b,\uparrow), (c,\downarrow), (d,\downarrow) \}$\\
      		$t_{(r1, r3)}$ & 1 & $\{(a,\downarrow), (b,\uparrow), (c,\downarrow), (d,\uparrow) \}$\\
      		$t_{(r1, r4)}, t_{(r3, r4)},$ & 3 & $\{(a,\downarrow), (b,\uparrow), (c,\uparrow), (d,\downarrow) \}$\\
      		$t_{(r2, r4)}$ & & \\
      		$t_{(r2, r3)}$ & 1 & $\{(a,\downarrow), (b,\uparrow), (c,\uparrow), (d,\uparrow) \}$\\
      		\bottomrule
    	\end{tabular}}%
    	\label{tab2:gp_encoding}
	\end{table}
		
	However, since GRAANK is based on a BFS strategy, it generates large numbers of candidates when dealing with data sets having huge numbers of attributes. This may lead to a combinatorial candidate explosion which overwhelms the algorithm. ParaMiner is based on DFS whose major drawback is finding the parent node of the tree with the longest length. DFS-based techniques employs recursion (which has an exponential computational complexity) to achieve this.
	
	In the section that follows, we propose a heuristic solution to the problem of combinatorial candidate explosion and finding parent node of an FP-Tree.
	
	\section{Related Works}
	\label{sec3:related}
	In this section, we present a review of some population-based optimization techniques that may be applied to the case of GP mining. In GP mining, evaluating all possible GP candidates in order to find those that are \textit{frequent} is computationally expensive, especially when a huge number of attributes is involved \cite{Di-Jorio2009,Fiot2009,Laurent2009,Owuor2019}.
	
	This problem may be alleviated by utilizing popula-tion-based optimization techniques (such as genetic algorithm (GA) \cite{Mirjalili2019}, particle swarm optimization (PSO) \cite{Kennedy1995} and ant colony optimization \cite{Dorigo1996}) to find the \textit{frequent} GP candidates without having to generate the whole candidate population.
	
	GA is among the first population-based optimization techniques to be proposed in history \cite{Goldberg1988,Mirjalili2019}. It is a meta-heuristic inspired by natural selection and the ``survival of the fittest'' systems originally proposed by Charles Darwin in the book titled ``The Origin of Species''. Based on the fitness value, in a competing environment, only the strongest individuals survive and attract mating partners. These individuals that are considered strong mate to produce offsprings and the good attributes from both parents may produce even stronger offsprings. In this way species become more well suited in their environment \cite{Goldberg1988,Holland1992,Mirjalili2019,Pourpanah2019}
	
	Data mining is one application area of GAs. For instance, Saravanan et al., \cite{Saravanan2014} propose a novel approach mining sequential patterns using GA. Kabir et al., \cite{Kabir2015} conduct a comparative study of GA-based techniques verses Apriori-based techniques for mining maximal frequent patterns and they prove that GA-based algorithms perform computationally better than their Apriori-based counterparts.
	
	PSO is a meta-heuristic algorithm implementation that (originally proposed by \cite{Kennedy1995} to imitate the movements flocks of birds or schools of fishes as they search for food) simulates the behaviors and movements of swarms in order to iteratively optimize a numeric problem \cite{Pourpanah2019,Rajamohana2018}. PSO can also be classified as an evolutionary computation technique that is inspired by the metaphor of social interaction and communication (such as bird flocking and fish schooling) \cite{Shruti2012}.
	
	In the realm of pattern mining, \cite{Shruti2011,Shruti2012} discover and prove that a PSO-based frequent pattern (FP)-growth technique is more efficient in mining fuzzy frequent patterns than traditional FP-growth techniques.
	
	ACO, as originally described by \cite{Dorigo1996}, is a general-purpose meta-heuristic approach for optimizing various combinatorial problems. It exploits the behavior of a colony of artificial ants in order to search for approximate solutions to discrete optimization problems \cite{Blum2005,Cicirello2001,Dorigo2019,Runkler2005,Silva2002}. The application areas for ACO are vast; for instance in the telecommunication domain, \cite{Sim2003} employed it to optimally load balance circuit-switched networks.
	
	It is important to highlight that GA and PSO based techniques use best solution and global best solution to generate new individual solutions \cite{Pourpanah2019}. On the contrary, ACO based techniques use pheromone levels to generate individual solutions \cite{Runkler2005}. In regards to GP mining, possible GP candidate solutions, more often, do not converge to a single best solution. Rather, candidate solutions are made up of combinations of 2 or more 1-item set GP solutions (see Section~\ref{sec2:preliminary}). Therefore, the problem is more of a `combinatorial problem' than a `fitness value' problem. In the sections that follow, we illustrate how ACO is better suited than GA and PSO in optimizing the problem of finding GP candidates.

	\section{Ant Colony Optimization for GP Extraction}
	\label{sec4:aco}
	
	\subsection{Ant Colony Optimization}
	\label{sec4.1:aco}
	ACO imitates the positive feedback reinforcement behavior of biological ants as they search for food \cite{Dorigo2010,Dorigo1996}. According to \cite{Hartman2008}, ACO utilizes a set of artificial ants to probabilistically contrive solutions $\mathcal{S}$ through a collective memory, \textit{pheromones} stored in matrix $\mathcal{T}$, together with a problem specific heuristic $\eta$.
	
	First, we introduce formal mathematical notations taken from literature concerning ACO. We use the traveling salesman problem (TSP)  \cite{Junger1995} in order to describe these preliminary mathematical notations \cite{Dorigo2010,Dorigo1996}.
	
	Given a set of $n$ towns, the TSP problem can be stated as the problem of finding the shortest route that visits each town once. The path between town $i$ and $j$ may be represented as $d_{i,j}$. Let $\langle b_{i}(t) \mid i = 1,2,...n \rangle$ be the number of ants in town $i$ at time $t$ and $q = \sum _{i=1} ^{n} b_{i}(t)$ be the total number of ants. Each ant is a simple agent that:
	\begin{itemize}
		\item chooses a town to go to based on the probability that is a function of the town distance and amount of pheromone present on the connecting edge;
		\item is prohibited from returning to already visited towns until a tour visit of all towns is completed; and
		\item deposits pheromone trails on each visited edge $(i,j)$, once it completes the tour.
	\end{itemize}
	
	Let $\tau _{i,j}(t)$ be the \textit{pheromone intensity} on edge $(i,j)$ at time $t$ and each ant at time $t$ chooses the next town to visit, where it will be at time $t+1$. We define $q$ movements made by $q$ ants in the interval $(t, t+1)$ a single \textit{iteration} of the ACO algorithm. Therefore, every $n$ iterations of the algorithm each ant completes a visit of all the towns (tour cycle) and at this point the pheromone intensity is updated according to the formula:
	
	\begin{equation}\label{eqn4:p_update}
		\tau _{i,j} (t+n) = (1-\rho) ~.~  \tau _{i,j} (t) +  \Delta \tau _{i,j}
	\end{equation}
	where $\rho$ (usually $<1$) is a coefficient such that $(1 - \rho)$ represents the evaporation of pheromone intensity between time $t$ and $t+n$,
	
	\begin{equation*}
		\Delta \tau _{i,j} = \sum _{k=1} ^{q} \Delta \tau _{i,j} ^{k}\notag
	\end{equation*}
	where $\Delta \tau _{i,j} ^{k}$ is the quantity per unit of length of phero-mone substance laid on edge $(i, j)$ by the $kth$ ant between time $t$ and $t+1$ and it is given by:
	
	\begin{equation*}
		 \Delta \tau _{i,j} ^{k} = 
		 \begin{cases}
		 	\frac{C}{L_{k}} \quad $ if $ kth $ ant uses path $ (i,j) \\ 
											\qquad \quad $(between time $ t $ and $ t+n)\\
			0 \qquad \; ~ otherwise
		 \end{cases}\notag
	\end{equation*}
	where $C$ is a constant and $L_{k}$ is the tour length of the $kth$ ant.
	
	In order to satisfy the constraint that an ant visits all the $n$ towns, each ant is associated with a \textit{tabu list} $(tabu_{k})$ that stores the towns already visited by the ant up to time $t$ and forbids it from re-visiting them until the \textit{tour cycle} is complete. At the end of the tour cycle, the tabu list is used to compute the distance of the path followed by the ant then, it is emptied and the next tour cycle begins.
	
	The probability $p_{i, j} ^{k} (t)$ of the $kth$ ant moving from town $i$ to $j$ as shown in Equation~\eqref{eqn4:rule_orig}:
	
	\begin{equation}\label{eqn4:rule_orig}
		p_{i, j} ^{k} (t) = 
		\begin{cases}
			\frac{[\tau _{i, j} (t)]^{\alpha} ~.~ [\eta _{i, j}]^{\beta}}{\sum _{k \in allow_{k}} [\tau _{i, k} (t)]^{\alpha} ~.~ [\eta _{i, k}]^{\beta}} \quad $if $ j\in allow_{k}\\
			0 \quad \qquad \qquad \qquad \qquad  \qquad \; otherwise
		\end{cases}
	\end{equation}
	where $\eta _{i, j} = 1/d_{i, j}$ and $allow_{k} = \{N - tabu_{k} \}$ and $\alpha$ and $\beta$ are parameters that control the relative importance of the pheromone intensity against visibility.
	
	
	In this paper, we will consider one variant of ACO called `\texttt{MAX-MIN} ant system' \cite{Stutzle2000} to optimize both BFS and DFS for the case of GP mining. The remainder of this section is organized as follows: we describe existing BFS and DFS traversal strategies for extracting gradual rules; we describe ACO approaches for optimizing these strategies and propose the corresponding mathematical notations.
	
	\subsection{ACO for BFS Candidate Generation}
	\label{sec4.2:aco_bfs}
	 
	In order to apply ACO to the GRAANK approach, we need to identify a suitable heuristic representation of the gradual item set candidate generation problem that can be solved using ACO.
	
	Given a set of $n = \{ a_{1},a_{2},... \}$ attributes of a data set $\mathcal{D}$ (as shown in Table~\ref{tab4:sample2}), GRAANK seeks to find \textit{frequent} GPs $M = \{ m_{1}, m_{2},...,m_{k} \}$ (where $\forall m \in M: m \subseteq n$) by generating and testing numerous candidates. 
	
	\clearpage
	
	\begin{table}[h!]
  		\centering
      	\caption{Sample data set $\mathcal{D}$ with attributes $\{a, b, c \}$.}
    	\begin{tabular}{l c c c}
      		\textbf{id} & \textbf{a} & \textbf{b} & \textbf{c}\\
      		\hline \hline
      		r1 & 5 & 30 & 43\\
      		r2 & 4 & 35 & 33\\
      		r3 & 3 & 40 & 42\\
      		r4 & 1 & 50 & 49\\
      		\bottomrule
    		\end{tabular}
    	\label{tab4:sample2}
	\end{table}
	
\begin{figure}[h!]
		\centering
		\small
  		\subfloat[]{%
		\begin{tikzpicture}
		\begin{scope}[circlestyle/.style={circle, thin, draw}]
			\node (nu) at (-0.6,1) {$a$};
			
			\node[circlestyle] (n11) at (-1.2,0) {$a+$};
			\node[circlestyle] (n12) at (0,0) {$a-$};

			\node[circlestyle] (n21) at (-0.2,-1.5) {$c+$};
			\node[circlestyle] (n22) at (0.8,-1.5) {$c-$};
			
			\node[circlestyle] (n31) at (-1.2,-3.2) {$b+$};
			\node[circlestyle] (n32) at (0,-3.2) {$b-$};
			
			\node (nd) at (-0.5,-4.2) {$b$};
		\end{scope}
	
		\begin{scope}
			\path (nu) edge[draw=black, dashed] (n11);
			\path (nu) edge[draw=black, dashed] (n12);
			
			\draw[-] (n11) edge (n31);
			\draw[-] (n11) edge (n21);
			\draw[-] (n12) edge (n22);
			\draw[-] (n21) edge (n31);
			\draw[-] (n21) edge (n32);
			
			\path (n31) edge[draw=black, dashed] (nd);
			\path (n32) edge[draw=black, dashed] (nd);
			
		\end{scope}
		\end{tikzpicture}
		}%
		$\qquad \qquad$
  		\subfloat[]{%
		\begin{tikzpicture}
		
		\begin{scope}[every node/.style={thin, draw}]
			\node (t) at (-1.7, 1.5) {t=0};
		\end{scope}
		
		\begin{scope}[circlestyle/.style={circle, thin, draw}]
			\node (nu) at (-0.6,1) {$a$};
			\node[above, blue] at (-0.6, 1.1) {\scriptsize 12 ants};
			
			\node[circlestyle] (n11) at (-1.2,0) {$a+$};
			\node[circlestyle] (n12) at (0,0) {$a-$};

			\node[circlestyle] (n21) at (-0.2,-1.5) {$c+$};
			\node[circlestyle] (n22) at (0.8,-1.5) {$c-$};
			
			\node[circlestyle] (n31) at (-1.2,-3.2) {$b+$};
			\node[circlestyle] (n32) at (0,-3.2) {$b-$};
			
			\node (nd) at (-0.5,-4.2) {$b$};
			\node[above, red] at (-0.5,-4.6) {\scriptsize 12 ants};
		\end{scope}
	
		\begin{scope}
			\path (nu) edge[draw=blue, dashed, ->] node[blue, sloped, above]{\tiny 6 ants} (n11);
			\path (nu) edge[draw=blue, dashed, ->] node[blue, sloped, above]{\tiny 6 ants} (n12);
			
			\draw[-] (n11) edge (n31);
			\draw[-] (n11) edge (n21);
			\draw[-] (n12) edge (n22);
			\draw[-] (n21) edge (n31);
			\draw[-] (n21) edge (n32);
			
			\path (nd) edge[draw=red, dashed, ->] node[red, sloped, below]{\tiny 6 ants} (n31);
			\path (nd) edge[draw=red, dashed, ->] node[red, sloped, below]{\tiny 6 ants} (n32);
			
			\draw[dashed, blue] ([xshift=-2ex]n11.south) edge[->]  node[sloped, below, blue] {\scriptsize 3 ants} ([yshift=10ex, xshift=-2ex]n31.north);
			\draw[dashed, blue] ([xshift=3ex]n11.south) edge[->]  node[sloped, above, blue] {\scriptsize 3 ants} ([yshift=0.6ex, xshift=-1ex]n21.north);
			\draw[dashed, blue] ([xshift=3ex]n12.south) edge[->]  node[sloped, above, blue] {\scriptsize 6 ants} ([yshift=0.6ex]n22.north);
			
			\draw[dashed, blue] ([xshift=-1ex, yshift=-1ex]n21.south) edge[->]  node[sloped, below, blue] {\tiny 1 ant} ([xshift=3ex]n31.north);

			\draw[dashed, red] ([xshift=-2ex]n31.north) edge[->]  node[sloped, above, red] {\scriptsize 3 ants} ([yshift=-10ex, xshift=-2ex]n11.south);
			\draw[dashed, red] ([yshift=2ex, xshift=2ex]n31.north) edge[->]  node[sloped, above, red] {\scriptsize 3 ants} ([xshift=-3ex]n21.south);
			\draw[dashed, red] ([yshift=1ex]n32.north) edge[->]  node[sloped, above, red] {\scriptsize 6 ants} ([xshift=1ex]n21.south);
			
		\end{scope}
				
		\end{tikzpicture}
		}%
		$\qquad$
  		\subfloat[]{%
		\begin{tikzpicture}
		
		\begin{scope}[every node/.style={thin, draw}]
			\node (t) at (-1.7, 1.5) {t=1};
		\end{scope}
		
		\begin{scope}[circlestyle/.style={cycle, thin, draw}]
			\node (nu) at (-0.6,1) {$a$};
			\node[above, blue] at (-0.6, 1.2) {\scriptsize 119 ants};
			
			\node (n11) at (-1.2,0) {$a+$};
			\node (n12) at (0.2,0) {$a-$};

			\node (n21) at (-0.2,-1.5) {$c+$};
			\node (n22) at (0.8,-1.5) {$c-$};
			
			\node (n31) at (-1.2,-3.2) {$b+$};
			\node (n32) at (0.2,-3.2) {$b-$};
			
			\node (nd) at (-0.5,-4.2) {$b$};
			\node[above, red] at (-0.5,-4.8) {\scriptsize 119 ants};
		\end{scope}
	
		\begin{scope}
			\path (nu) edge[draw=blue, dashed, ->] node[blue, sloped, above]{\tiny 65 ants} (n11);
			\path (nu) edge[draw=blue, dashed, ->] node[blue, sloped, above]{\tiny 34 ants} (n12);
			
			\draw[thick] (n11) edge (n31);
			\draw[very thick] (n11) edge (n21);
			\draw[thick] (n12) edge (n22);
			\draw[] (n21) edge (n31);
			\draw[very thick] (n21) edge (n32);
			
			\path (nd) edge[draw=red, dashed, ->] node[red, sloped, below]{\tiny 63 ants} (n31);
			\path (nd) edge[draw=red, dashed, ->] node[red, sloped, below]{\tiny 56 ants} (n32);
			
			\draw[dashed, blue] ([xshift=-2ex]n11.south) edge[->]  node[sloped, below, blue] {\scriptsize 26 ants} ([yshift=10ex, xshift=-2ex]n31.north);
			\draw[dashed, blue] ([xshift=2ex]n11.south) edge[->]  node[sloped, above, blue] {\scriptsize 39 ants} ([yshift=0.5ex]n21.north);
			\draw[dashed, blue] ([xshift=2ex]n12.south) edge[->]  node[sloped, above, blue] {\scriptsize 56 ants} ([yshift=1.5ex]n22.north);
			
			\draw[dashed, blue] ([yshift=1ex]n21.south) edge[->]  node[sloped, below, blue] {\tiny 13 ants} ([xshift=2ex]n31.north);

			\draw[dashed, red] ([xshift=-2ex]n31.north) edge[->]  node[sloped, above, red] {\scriptsize 42 ants} ([yshift=-10ex, xshift=-2ex]n11.south);
			\draw[dashed, red] ([yshift=0ex]n31.north) edge[->]  node[sloped, above, near end, red] {\tiny 21 ants} ([xshift=-2ex]n21.south);
			\draw[dashed, red] ([yshift=1ex]n32.north) edge[->]  node[sloped, above, red] {\scriptsize 56 ants} ([xshift=1.5ex]n21.south);
			
		\end{scope}		
		\end{tikzpicture}
		}%
    	\caption{\footnotesize An example of artificial ants for BFS GP mining: (a) initial paths, (b) at time $t=0$ there is no pheromone intensity on any path; so ants choose paths with equal probability, and (c) at time $t=1$ the pheromone intensity is stronger on paths with cheapest gradual variations; therefore, ants prefer these paths.}
  	\label{fig4:artificial_ants_bfs}
  	\end{figure}
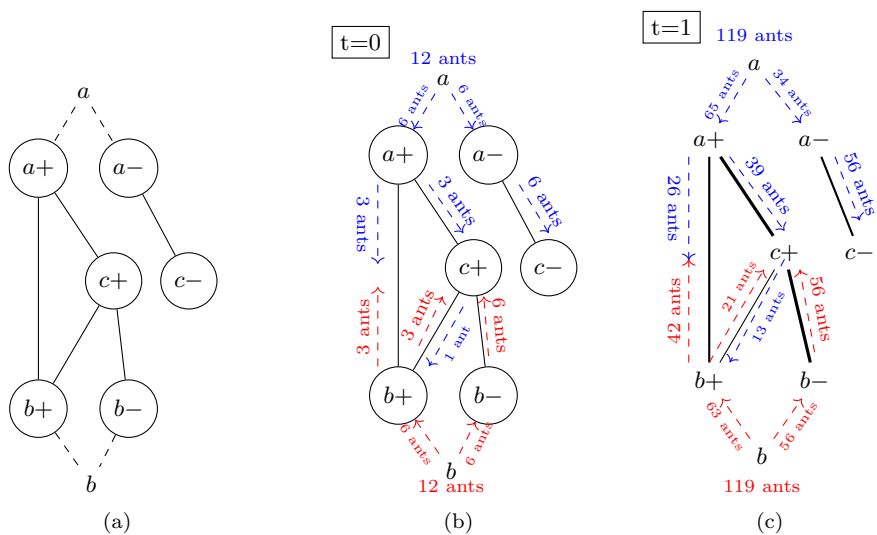

	\textit{Example 4.1.} We consider a sample graphs of artificial ants moving on the edges of gradual items $a+, a-,$\\$ b+, b-, c+$ and $c-$ (where $+$ denotes `increasing', $-$ denotes `decreasing', $a+$ implies $(a, \uparrow)$, $a-$ implies $(a, \downarrow)$, and minimum support threshold $\sigma = 0.5$) as shown in Figure~\ref{fig4:artificial_ants_bfs}. For the sake of simplicity, we remove all edges connecting to nodes (or gradual items) whose \textit{frequency support} ($sup$)$< \sigma$ and assume that $sup(\{a+, c+ \}) > \sigma$ but $sup(\{a+, c+, b+ \}) < \sigma$.

	In order to make an accurate interpretation of the ACO for BFS, assume that the distances of all node edges are equal. Let us consider what happens at regular discrete time intervals: $t=0,1,2,...$ . Suppose that 12 new ants come to nodes $a$ and $b$ respectively at time $t=0$ and another 119 new ants to nodes $a$ and $b$ at time $t=1,2,...$ . Each ant travels at a speed of 2 \textit{edge distances per time interval} and that by moving along at time $t$ it deposits pheromones of intensity 1 only if all the nodes visited up to that time $t$ form a gradual item set whose $support \geq \sigma$. The pheromone evaporates by a factor $(1 - \rho)$ between successive time interval $(t+1, t+2)$.

	At time $t=0$ no path has any pheromone intensity; therefore, the 12 news ants at nodes $a$ and $b$ choose all paths  with equal probability as shown in Figure~\ref{fig4:artificial_ants_bfs}b. Since $sup(\{a+, c+, b+ \}) < \sigma$ the ant from $a$ that chose edges between nodes $a+$, $c+$ and $b+$ and the 3 ants from $b$ that chose edges between nodes $b+$, $c+$ and $c+$ do not deposit any pheromone on any of these edges.
	
	At time $t=1$ 119 new ants arrive at nodes $a$ and $b$. The 119 new ants at node $a$ will find a pheromone intensity 6 on the path that leads to $a-$ deposited by ants that used it previously from node $a-$ and a pheromone intensity of 15 at node $a+$: 6 on the path that leads to $b+$ (3 deposited by ants that departed from $a+$ and 3 deposited by ants that arrived from $b+$), 9 on the path that leads to $c+$ (3 deposited by ants that departed from $a+$ and 6 deposited by ants that arrived from $c+$). The probability of the ants choosing to move towards node $a-$ is 6/21, that of moving towards node $a+$ is 15/21, that of moving towards node $b+$ is 6/15 and that of moving towards $c+$ 9/15. The same is true for the new 119 ants at node $b$, as shown in Figure~\ref{fig4:artificial_ants_bfs}c. Therefore, the best variation routes (or candidates) are: $\{a+,c+,b-\},\{a+,b+\},\{a-,c-\}$.

	\subsubsection{Proposed Mathematical Notations of ACO for GPCG Problem}
	\label{sec4.2.1:aco_bfs_notations}
	
	The gradual pattern candidate generation (GPCG) pro-blem can be stated as: 
	
	\textit{``the problem of finding the cheapest variation routes that connect all gradual items that satisfy the minimum frequency requirement once.''}
	
	In GP mining, a potential candidate must not include any \textit{infrequent} singleton gradual item set (anti-monotonicity property \cite{Owuor2019}). Therefore, when apply ACO to the case of GPCG problem, we define each node of the ACO graph as a singleton gradual item derived from its attribute and the variation edge distance $d_{i, j}$ between nodes $i, j$ is given by formula in Equation~\eqref{eqn4:bfs_distance}:
	
	\begin{equation}\label{eqn4:bfs_distance}
		d_{i, j} = 
		 \begin{cases}
		 	1 \qquad \quad $if $ sup(i) \geq \sigma $ and $ sup(j) \geq \sigma\\
			\infty \qquad \; otherwise
		 \end{cases}
	\end{equation}
	where $sup(i)$ is the frequency support of gradual item set $\{i\}$, $\sigma$ is a user-specified minimum support threshold - see Section~\ref{sec2.2:gps}. The possibility of gradual item $(i, j)$ being frequent is either visible `1' or not visible `$\infty$'.
	
	Let $\langle b_{i}(t) \mid i = 1,2,...n \rangle$ be the number of ants at node $i$ at time $t$ and $q = \sum _{i=1} ^{n} b_{i}(t)$ be the total number of ants. Each ant is a simple agent that:
	\begin{itemize}
		\item chooses a node to go to based on the probability that is a function of the amount of pheromone present on the connecting edge;
		\item is prohibited from returning to already visited nodes until a tour visit of all nodes is completed (this is managed by a \textit{tabu list}); and
		\item deposits pheromone trails on all visited edges if support of all visited nodes $sup(\mathbf{N}) \geq \sigma$, once it completes the tour. This is because a gradual candidate is formed by combining all the visited nodes into a set; and if the candidate is not frequent then its path is also not appealing. Lastly, every edge $(i, j)$ is removed if: $sup(i) < \sigma$ and $sup(j) < \sigma$.
	\end{itemize}

	Let $\tau _{i,j}(t)$ be the \textit{pheromone intensity} on edge $(i,j)$ at time $t$ and each ant at time $t$ chooses the next node to visit, where it will be at time $t+1$. We define $q$ movements made by $q$ ants in the interval $(t, t+1)$ a single \textit{iteration} of the ACO algorithm. Therefore, every $n$ iterations of the algorithm each ant completes a visit of all valid nodes (tour cycle) and at this point the pheromone intensity is updated according to the formula in Equation~\eqref{eqn4:p_update_bfs}.
	
	\begin{equation}\label{eqn4:p_update_bfs}
		 \tau _{i, j} (t+n) = 
		 \begin{cases}
		 	 (1-\rho) . \tau _{i, j} (t) +  \Delta \tau _{i, j} \quad $if $ sup(\mathbf{N}_{n}) \geq \sigma\\
			0 \qquad \qquad \qquad \qquad \qquad otherwise
		 \end{cases}
	\end{equation}
	where $\rho$ and $\Delta \tau _{i, j}$ are similar to Equation~\eqref{eqn4:p_update} and $\mathbf{N}_{n}$ is the set of all nodes visited by the ant at end of $n$ iterations
	
	\begin{equation*}
		\mathbf{N}_{n} = \{i_{t}\}_{t=0} ^{t=n}\notag
	\end{equation*}
	
	Similarly, each ant is associated with a tabu list $tabu_{k}$ that stores the nodes already visited up to time $t$ and forbids the ant from revisiting them until the tour cycle is complete. In addition, the tabu list also keeps track of node edges which do not meet the minimum frequency support requirement. Since we modify notation $d_{i, j}$ in Equation~\eqref{eqn4:bfs_distance}, visibility ($\eta _{i, j} = 1/d_{i, j}$) is consequently affected. The probability $p_{i, j} ^{k} (t)$ of the $kth$ ant moving from node $i$ to $j$ becomes:
	
	\begin{equation}\label{eqn4:rule_bfs}
		p_{i, j} ^{k} (t) = 
		\begin{cases}
			\frac{[\tau _{i, j} (t)]^{\alpha}}{\sum _{k \in allow_{k}} [\tau _{i, k} (t)]^{\alpha}} \quad $if $ j\in allow_{k} $ and $\\
			\qquad \qquad \qquad \qquad \quad (sup(i),sup(j) \geq \sigma)\\
			0 \qquad \qquad \qquad \qquad otherwise
		\end{cases}
	\end{equation}
	\begin{center}
	where $allow_{k}$ and $\alpha$ are similar to Equation~\eqref{eqn4:rule_orig}.
	\end{center}
	
	Therefore, the probability $p_{i, j} ^{k} (t)$ is a function of pheromone intensity at time $t$: $\tau _{i, j} (t)$ (which instructs that the edge $(i, j)$ with most ant traffic should be selected with highest probability - implementing an autocatalytic process). In GP mining \textit{visibility} between nodes $i$ and $j$ can be compared to the possibility of forming a candidate by combining node $i$ and $j$; and it is either `visible' to the ant if both nodes are \textit{frequent} or `invisible' to the ant if one of the nodes is \textit{infrequent}. So, $\eta _{i, j}$ is denoted as either 1 or 0 for `visible' or `invisible' respectively.
	
	\subsubsection{ACO-GRAANK Algorithm}
	\label{sec4.2.2:aco-graank}
	
	ACO-GRAANK algorithm is an optimized version of the GRAANK algorithm \cite{Laurent2009} and it inherits the efficient binary representation of concordant tuple pairs, but applies ACO to generate candidate item sets.
	
	We take the position that a gradual item set may also be referred to as a \textbf{pattern solution}. In that case all possible gradual item set solutions ($\mathcal{S}_{n}$) are admissible and can be generated based on the pheromone matrix ($\mathcal{T}_{a,j}$). Notwithstanding, all the gradual item sets in a generated solution will be evaluated and the solution updated with only valid item sets.

	First we mention that initially there exists an equal chance for any attribute $A$ (of data set $\mathcal{D}$) to either increase ($+$) or decrease ($-$) or be irrelevant ($\times$). As the algorithm acquires more knowledge about valid patterns, the possibilities of the 3 options are adjusted accordingly. We propose an artificial pheromone matrix of 2-Dimensional whose size is shown in Equation~\eqref{eqn4:p-matrix_graank}. The matrix contains knowledge about pheromone proportions of the 3 gradual options for each attribute.
		
	\begin{equation}
	\label{eqn4:p-matrix_graank}
		\mathcal{T}_{a, j} = q \times 3
	\end{equation}
	\begin{center}
		where $q$: number of attributes and $j\in \{+, -, \times\}$
	\end{center}
	
	At the beginning all artificial pheromones $p_{a, j}$ in matrix $\mathcal{T}_{a, j}$ are initialized to 1, then they are updated as follows at time interval $(t, t+1)$:
	\begin{itemize}
		\item every generated gradual item set solution is evaluated and only valid solutions are used to update the artificial pheromone matrix. Invalid solutions are stored and used to reject their super-sets; and
		\item in a given valid solution, each gradual item set is used to update the corresponding pheromone $p_{a, j}(t)$ (where $j$ is either $+$ or $-$) by formula: $p_{a, j}(t) = p_{a, j}(t) + 1$.
	\end{itemize}
	
	The probability rule $\mathcal{P}_{a, j} (t)$ is given by calculating the proportions of its artificial pheromones $p_{a*}$, as shown in Equation~\eqref{eqn4:acograank_rule}.
	
	\begin{equation}\label{eqn4:acograank_rule}
		\mathcal{P}_{a, j} (t) = \frac{p_{a, j}}{\sum _{k \in \{+, -, \times \}} p_{a, k} (t)}
	\end{equation}

	The main steps of ACO-GRAANK algorithm are:
	
	\begin{enumerate}
		\item Build binary matrices of 1-itemset from input data set $\mathcal{D}$ (described in GRAANK \cite{Laurent2009}).
		\item Generate gradual item sets from pheromone matrix $\tau_{aj}$ and validate each generated item set by comparing its support against the minimum support $\sigma$.
		\item If the generated pattern is valid, use gradual items to update the pheromone matrix.
		\item Repeat steps 2-3 until the algorithm starts to generate similar gradual item sets.
	\end{enumerate}
	
	\subsubsection{Computational Complexity of ACO-GRAANK}
	\label{sec4.2.3}
	The big-O notation \cite{Cormen2009,Pourpanah2019,Vaz2017} is used to analyze the computational complexity of ACO-GRAANK. For every GP candidate that is generated, ACO-GRAANK algorithm builds binary matrices, performs a binary AND operation on the matrices and calculates support \cite{Laurent2009}. We formulate the problem and show its complexity.
	
	\textbf{Problem formulation.} Given a dataset $\mathcal{D}$ with $m$ attributes and $n$ objects, we can generate numerous GP candidates each with $k$ item sets (where $k \leq m$). For example, if $\mathcal{D}$ has 3 attributes $\{a, b, c\}$, one of the candidates may be $\{(a,\uparrow), (b,\uparrow)\}$ which has 2 item sets. For each candidate, we build binary matrices of every item set as shown in Table~\ref{tab2:binary_matrices}. Next we perform a binary AND on these binary matrices and calculate frequency support. Using big-O notation building binary matrices results in a complexity of $O(k\cdot n^{2})$. The binary AND operation and support calculation have small complexities in comparison to that of the latter operation.
	
	\textbf{Search space size.} It is important to highlight that, GP candidate search space grows almost exponentially as the number of attributes increase. Given a huge search space, the aim of ACO-GRAANK algorithm is to find useful candidates (candidates whose frequency support $\geq \sigma$) and ignore useless candidates in the search space. This can be achieved by minimizing the number of algorithm iterations required to learn the valid candidates. Therefore, if an ACO-GRAANK algorithm iterates $x$ times, it generates $x$ GP candidates and its computational complexity is $O(x\cdot k\cdot n^{2})$.

	\subsection{ACO for DFS FP-Tree Search}
	\label{sec4.3:aco_dfs}
	Similarly, for the purpose of applying ACO to a DFS-based approach, we identify a suitable heuristic representation to the problem of finding the parent node of the longest FP-tree.
	
	Given a set of $n=\{a_{1}, a_{2}, ..., a_{k}\}$ attributes of a data set where each attribute has a set of tuples $a_{k} = \{r_{1}, r_{2}, ... \}$ (as shown in Table~\ref{tab4:sample3}a), gradual DFS techniques seek to find \textit{frequent} GPs $M = \{ m_{1}$, $m_{2},... \}$ (where $\forall m \in M: m \subseteq n$) by encoding the data set into a transactional data set (as shown in Table~\ref{tab4:sample3}b) and recursively searching the transactional data set for the longest FP-Tree.
	
	\begin{table}[h!]
		\centering
		\caption{(a) Sample data set $\mathcal{D}$ and (b) its corresponding sorted reduced transactional data set with minimum length of \texttt{tid} is 3 - see Section~\ref{sec2.3:gradual_traversal}.}
		\subfloat[]{%
		\begin{tabular}{l c c c c}
      		\textbf{id} & \textbf{a} & \textbf{b} & \textbf{c} & \textbf{d} \\
      		\hline \hline
      		r1 & 5 & 30 & 43 & 97\\
      		r2 & 4 & 35 & 33 & 86\\
      		r3 & 3 & 40 & 42 & 108\\
      		r4 & 1 & 50 & 49 & 27\\
      		\bottomrule
    	\end{tabular}}%
			\\
		\subfloat[]{%
		\begin{tabular}{c l}
      		\textbf{item} & \textbf{tids} \\
      		\hline \hline
      		$(a,\downarrow)$ & $\{t_{(r1, r2)}, t_{(r1, r3)}, t_{(r1, r4)},$\\
          	& $t_{(r2, r3)}, t_{(r2, r4)}, t_{(r3, r4)} \}$\\
			$(b,\uparrow)$ & $\{t_{(r1, r2)}, t_{(r1, r3)}, t_{(r1, r4)},$\\
          	& $t_{(r2, r3)}, t_{(r2, r4)}, t_{(r3, r4)} \}$\\   
      		$(c,\uparrow)$ & $\{t_{(r1, r4)}, t_{(r2, r3)}, t_{(r2, r4)},$\\
      		& $t_{(r3, r4)} \}$\\
      		$(d,\downarrow)$ & $\{t_{(r1, r2)}, t_{(r1, r4)}, t_{(r2, r4)},$\\
          	& $t_{(r3, r4)} \}$\\
      		\bottomrule
    	\end{tabular}
		}%
    	\label{tab4:sample3}
	\end{table}

\begin{figure}[h!]
		\centering
		\small
  		\subfloat[]{%
		\begin{tikzpicture}
		\begin{scope}
			\node (n1) at (0,1) {};
			\node (n2) at (0,0) {$r_{1}$};
			\node (n3) at (0.8,-1) {$r_{4}$};
			\node (n4) at (-1.5,-1) {$r_{3}$};
			\node (n5) at (0,-2) {$r_{2}$};
			\node (n6) at (0,-3) {};
		\end{scope}
	
		\begin{scope}
			\path (n1) edge[draw=black, dashed] (n2);
			\draw[-] (n2) edge  node[sloped, above, brown] {\scriptsize d=0.25} (n3);
			\draw[-] (n2) edge  node[sloped, above, brown] {\scriptsize d=0.5} (n4);
			\draw[-] (n3) edge  node[sloped, below, brown] {\scriptsize d=0.25} (n5);
			\draw[-] (n4) edge  node[sloped, below, brown] {\scriptsize d=0.33} (n5);
			\path (n5) edge[draw=black, dashed] (n6);
		\end{scope}
		\end{tikzpicture}
		}%
		$\qquad \qquad$
  		\subfloat[]{%
		\begin{tikzpicture}
		
		\begin{scope}[every node/.style={thin, draw}]
			\node (t) at (-1.2, 1.5) {t=0};
		\end{scope}
		
		\begin{scope}
			\node (n1) at (0,1) {};
			\node (n2) at (0,0) {$r_{1}$};
			\node (n3) at (0.8,-1) {$r_{4}$};
			\node (n4) at (-1.5,-1) {$r_{3}$};
			\node (n5) at (0,-2) {$r_{2}$};
			\node (n6) at (0,-3) {};
		\end{scope}
			
		\begin{scope}
			\path (n2) edge[draw=black] (n3);
			\path (n2) edge[draw=black] (n4);
			\path (n3) edge[draw=black] (n5);
			\path (n4) edge[draw=black] (n5);

			\draw[blue] (n2) edge[dashed, <-]  node[sloped, below, blue] {\tiny 12 ants} (n1);
			\draw[dashed, blue] ([xshift=2.5ex]n2.south) edge[->]  node[sloped, above, blue] {\scriptsize 6 ants} (n3.north);
			\draw[dashed, blue] ([xshift=-4ex]n2.south) edge[->]  node[sloped, above, blue] {\scriptsize 6 ants} ([xshift=1ex]n4.north);
			
			\draw[red] (n6) edge[dashed, ->]  node[sloped, below, red] {\tiny 12 ants} (n5);
			\draw[dashed, red] ([xshift=2.5ex]n5.north) edge[->]  node[sloped, below, red] {\scriptsize 6 ants} ([yshift=0.2ex]n3.south);
			\draw[dashed, red] ([xshift=-4ex]n5.north) edge[->]  node[sloped, below, red] {\scriptsize 6 ants} ([xshift=1ex]n4.south);
		\end{scope}
		
		\end{tikzpicture}
		}%
		$\qquad \qquad$
  		\subfloat[]{%
		\begin{tikzpicture}
		
		\begin{scope}[every node/.style={thin, draw}]
			\node (t) at (-1.2, 1.5) {t=1};
		\end{scope}
		
		\begin{scope}
			\node (n1) at (0,1) {};
			\node (n2) at (0,0) {$r_{1}$};
			\node (n3) at (0.8,-1) {$r_{4}$};
			\node (n4) at (-1.5,-1) {$r_{3}$};
			\node (n5) at (0,-2) {$r_{2}$};
			\node (n6) at (0,-3) {};
		\end{scope}
	
		\begin{scope}
			\path (n2) edge[draw=black, very thick] (n3);
			\path (n2) edge[draw=gray, very thin] (n4);
			\path (n3) edge[draw=black, very thick] (n5);
			\path (n4) edge[draw=gray, very thin] (n5);

			\draw[blue] (n2) edge[dashed, <-]  node[sloped, below, blue] {\tiny 12 ants} (n1);
			\draw[dashed, blue] ([xshift=2.5ex]n2.south) edge[->]  node[sloped, above, blue] {\scriptsize 8 ants} (n3.north);
			\draw[dashed, blue] ([xshift=-4ex]n2.south) edge[->]  node[sloped, above, blue] {\scriptsize 4 ants} ([xshift=1ex]n4.north);
			
			\draw[red] (n6) edge[dashed, ->]  node[sloped, below, red] {\tiny 12 ants} (n5);
			\draw[dashed, red] ([xshift=2.5ex]n5.north) edge[->]  node[sloped, below, red] {\scriptsize 8 ants} ([yshift=0.2ex]n3.south);
			\draw[dashed, red] ([xshift=-4ex]n5.north) edge[->]  node[sloped, below, red] {\scriptsize 4 ants} ([xshift=1ex]n4.south);
		\end{scope}
		
		\end{tikzpicture}
		}%
    	\caption{\footnotesize An example of artificial ants for DFS: (a) initial paths with distances, (b) at time $t=0$ there is no pheromone intensity on any path; so ants choose all paths with equal probability and (c) at time $t=1$ the pheromone intensity is stronger on the shorter paths; therefore more ants prefer these paths.}
  	\label{fig4:artificial_ants_dfs}
  	\end{figure}
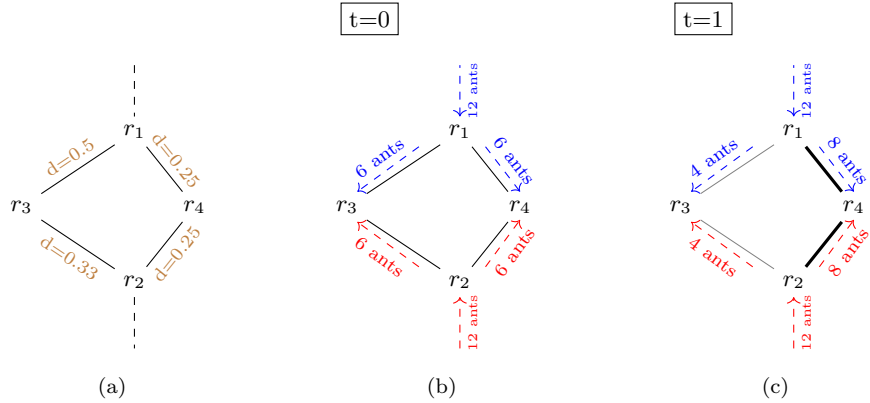

	\textit{Example 4.2.} We consider a sample graphs of artificial ants moving on the edges of nodes (or tuples) $r_{1}, r_{2}, r_{3}$ and $r_{4}$ as shown in Figure~\ref{fig4:artificial_ants_dfs}. For the case of DFS in GP mining, we propose to use the occurrence count of tuples in the encoded transactional data set to determine the length of distance between nodes (such as $d_{i, j} = \frac{1}{1 + \sum (r_{i}, r_{j})_{count}}$).
			
  	In order to make an accurate interpretation of ACO for BFS, assume that the paths between nodes $r_{1}$ and $r_{3}$, $r_{3}$ and $r_{2}$ have longer lengths than paths between nodes $r_{1}$ and $r_{4}$, $r_{4}$ and $r_{2}$ (as indicated by the distances in Figure~\ref{fig4:artificial_ants_dfs}a). Let us consider what happens at regular discrete time intervals: $t=0,1,2...$ . Suppose that 12 new ants come to node $r_{1}$ and 12 new ants come to node $r_{2}$ at each time interval. Each ant travels at a speed of 0.5 \textit{distance per time interval} and that by moving along at time $t$ it deposits pheromones of intensity 1, which completely and instantaneously evaporates in the middle of the successive time interval $(t+1, t+2)$.
  	
  	\clearpage
  	
	At time $t=0$ no path has any pheromone intensity and there are have 12 new ants at node $r_{1}$ and 12 new ants at node $r_{2}$. The ants select between nodes $r_{4}$ and $r_{3}$ randomly; therefore, on average 6 ants from each node move towards $r_{4}$ and 6 ants towards $r_{3}$ - as shown in Figure~\ref{fig4:artificial_ants_dfs}b.
	
	At time $t=1$ there are 12 new ants at node $r_{1}$ and 12 new ants at node $r_{2}$. The 12 new ants at node $r_{1}$ will find a pheromone intensity of 6 on the path that leads to node $r_{3}$ deposited by the 6 ants that used it previously from node $r_{1}$ and a pheromone intensity of 12 on the path that leads to node $r_{4}$ deposited by 6 ants that used it coming from node $r_{1}$ and 6 that arrived from node $r_{2}$. Therefore, the ants will find the path to node $r_{4}$ more desirable than that to node $r_{3}$. The probability of the ants choosing to move towards node $r_{4}$ is 2/3, while that of moving towards node $r_{3}$ is 1/3 and 8 ants will move towards node $r_{4}$ and 4 ants towards node $r_{3}$. The same is true for the 12 new ants at node $r_{2}$, as shown in Figure~\ref{fig4:artificial_ants_dfs}c. Therefore, the most appealing FP-Tree is $\{(r_{1}, r_{4}), (r_{2}, r_{4})\}$ which both appear in transactions with gradual items $\{ (b,\uparrow), (c,\uparrow), (d,\downarrow) \}$
	
	\subsubsection{Proposed Mathematical Notations of ACO for GPFP-Tree Problem}
	\label{sec4.3.1:aco_dfs_notations}
	
	The gradual pattern FP-Tree (GPFP-Tree) problem can be stated as: 
		
	\textit{``the problem of finding the longest FP-Tree from which frequent gradual patterns can be constructed.''}
	
	In the case of DFS for GP mining, we inherit the concept of encoding data sets into transactional data sets and harness this to represent the GPFP-Tree problem as a slightly different version of the TSP problem. Given a set of $n$ tuples of a data set, we seek to find the cheapest route that visits the most tuples once. The path between tuples $i$ and $j$ may be represented as $d_{i, j}$ and an instance of the GPFP-Tree may also be represented as a graph composed of tuples (or nodes) and edges. Each edge distance is derived from the formula:
	
	\begin{equation}\label{eqn4:dfs_distance}
		d_{i, j} = \frac{1}{1 + \sum (r_{i}, r_{j})_{count}}
	\end{equation}
	where $\sum (r_{i}, r_{j})_{count}$ is the occurrence count of tuple pair $(r_{i}, r_{j})$ in an encoded transactional data set.
	
	Let $\langle b_{i}(t) \mid i = 1,2,...n \rangle$ be the number of ants in node $i$ at time $t$ and $q = \sum _{i=1} ^{n} b_{i}(t)$ be the total number of ants. Each ant is a simple agent that:
	\begin{itemize}
		\item chooses a node to go to based on the probability that is a function of the node distance and amount of pheromone present on the connecting edge;
		\item is prohibited from returning to already visited nodes until a tour visit of all allowed nodes is completed (this is managed by a \textit{tabu list}); and
		\item deposits pheromone trails on each visited edge $(i, j)$, once it completes the tour.
	\end{itemize}
	
	Let $\tau _{i, j}(t)$ be the \textit{pheromone intensity} on edge $(i, j)$ at time $t$ and each ant at time $t$ chooses the next node to visit, where it will be at time $t+1$. We define $q$ movements made by $q$ ants in the interval $(t, t+1)$ a single \textit{iteration} of the ACO algorithm. Therefore, every $n$ iterations of the algorithm each ant completes a visit of all the nodes (tour cycle) and at this point the pheromone intensity is updated according to the formula in Equation~\eqref{eqn4:p_update_dfs}.
	
	\begin{equation}\label{eqn4:p_update_dfs}
		\tau _{i, j} (t+n) = (1-\rho) ~.~  \tau _{i, j} (t) +  \Delta \tau _{i, j}
	\end{equation}
	\begin{center}
		where $\rho$ and $\Delta \tau _{i, j}$ are similar to Equation~\eqref{eqn4:p_update}.
	\end{center}
	
	Similar to the original ACO, each ant is associated with a \textit{tabu list} $(tabu_{k})$ that stores the nodes already visited by the ant up to time $t$ and forbids it from re-visiting them until the \textit{tour cycle} is complete. At the end of the tour cycle, the tabu list is used to compute the distance of the path followed by the ant then, it is emptied and the next tour cycle begins. The probability $p_{i, j} ^{k} (t)$ of the $kth$ ant moving from node $i$ to $j$ as shown in Equation~\eqref{eqn4:rule_dfs}.
	
	\begin{equation}\label{eqn4:rule_dfs}
		p_{i, j} ^{k} (t) = 
		\begin{cases}
			\frac{[\tau _{i, j} (t)]^{\alpha} ~.~ [\eta _{i, j}]^{\beta}}{\sum _{k \in allow_{k}} [\tau _{i, k} (t)]^{\alpha} ~.~ [\eta _{i, k}]^{\beta}} \qquad $if $ j\in allow_{k}\\
			0 \qquad \qquad \qquad \qquad \qquad  \quad \; ~ otherwise
		\end{cases}
	\end{equation}
	\begin{center}
		where $allow_{k}$ and $\alpha$ and $\beta$ are similar to Equation~\eqref{eqn4:rule_orig}.
	\end{center}
	
	Therefore, in the GPFP-Tree problem probability $p_{i, j} ^{k} (t)$ is a trade-off between $\eta _{i, j}$ and $\tau _{i, j} (t)$. Put differently, it a trade-off between a greedy constructive heuristic and an autocatalytic process.
	
	\subsubsection{ACO-ParaMiner Algorithm}
	\label{sec4.3.2:aco-paraminer}
	
	The ACO-ParaMiner algorithm which is an optimized version of ParaMiner technique proposed by \cite{Negrevergne2014}. ACO-ParaMiner inherits the technique of transactional encoding from ParaMiner, but applies ACO to find the longest FP-Tree. 

	We propose the following ACO optimizations to the ParaMiner approach (proposed in \cite{Negrevergne2014}): (1) in addition to reducing the transactional data set by combining similar item set transactions and removing infrequent ones, we construct a \textbf{cost matrix} of all corresponding nodes. (2) we replace the recursive function with a non-recursive heuristic function that quickly learns the longest FP-tree with the help of the \textbf{cost matrix}.
		
	To illustrate, we use data set $\mathcal{D}$ in Table~\ref{tab4:sample4}. In order to remove infrequent items, the transactional data set is sorted by item occurrence as shown in Table~\ref{tab4:transactional2}b. If we set the minimum length of \texttt{tids} to 3, we remove infrequent items as illustrated in Table~\ref{tab4:trans_costmatrix}a.
	
	\begin{table}[h!]
  		\centering
      	\caption{Sample data set $\mathcal{D}$.}
    	\begin{tabular}{l c c c c}
      		\textbf{id} & \textbf{a} & \textbf{b} & \textbf{c} & \textbf{d} \\
      		\hline \hline
      		r1 & 5 & 30 & 43 & 97\\
      		r2 & 4 & 35 & 33 & 86\\
      		r3 & 3 & 40 & 42 & 108\\
      		r4 & 1 & 50 & 49 & 27\\
      		\bottomrule
    		\end{tabular}
    	\label{tab4:sample4}
	\end{table}

	\begin{table}[h!]
		\centering
		\caption{(a) example of a reduced transactional data set, (b) sorted items by occurrence.}
		\subfloat[]{%
		\begin{tabular}{l c l}
      		\textbf{tids} & \textbf{w} & \textbf{item-sets} \\
      		\hline \hline
      		$t_{(r1, r2)}$ & 1 & $\{(a,\downarrow), (b,\uparrow), (c,\downarrow), (d,\downarrow) \}$\\
      		$t_{(r1, r3)}$ & 1 & $\{(a,\downarrow), (b,\uparrow), (c,\downarrow), (d,\uparrow) \}$\\
      		$t_{(r1, r4)}, t_{(r2, r4)}$ & 3 & $\{(a,\downarrow), (b,\uparrow), (c,\uparrow), (d,\downarrow) \}$\\
      		$t_{(r3, r4)}$ & &\\
      		$t_{(r2, r3)}$ & 1 & $\{(a,\downarrow), (b,\uparrow), (c,\uparrow), (d,\uparrow) \}$\\
      		\bottomrule
    	\end{tabular}}%
			\\
		\subfloat[]{%
		\begin{tabular}{c l}
      		\textbf{item} & \textbf{tids} \\
      		\hline \hline
      		$(a,\downarrow)$ & $\{t_{(r1, r2)}, t_{(r1, r3)}, t_{(r1, r4)}, t_{(r2, r3)},$\\
          	& $t_{(r2, r4)}, t_{(r3, r4)} \}$\\
			$(b,\uparrow)$ & $\{t_{(r1, r2)}, t_{(r1, r3)}, t_{(r1, r4)}, t_{(r2, r3)},$\\
          	& $t_{(r2, r4)}, t_{(r3, r4)} \}$\\   
      		$(c,\uparrow)$ & $\{t_{(r1, r4)}, t_{(r2, r3)}, t_{(r2, r4)}, t_{(r3, r4)} \}$\\
      		$(d,\downarrow)$ & $\{t_{(r1, r2)}, t_{(r1, r4)}, t_{(r2, r4)}, t_{(r3, r4)} \}$\\
      		$(c,\downarrow)$ & $\{t_{(r1, r2)}, t_{(r1, r3)} \}$\\
      		$(d,\uparrow)$ & $\{t_{(r1, r3)}, t_{(r2, r3)} \}$\\  
      		$(a,\uparrow)$ & $\{\emptyset \}$\\
         	$(b,\downarrow)$ & $\{\emptyset \}$\\
      		\bottomrule
    	\end{tabular}
			}%
    	\label{tab4:transactional2}
	\end{table}
	
	\begin{table}[h!]
		\centering
		\caption{(a) sorted reduced transactional data set, (b) corresponding \textbf{cost matrix}.}
		\subfloat[]{%
		\begin{tabular}{c l}
      		\textbf{item} & \textbf{tids} \\
      		\hline \hline
      		$(a,\downarrow)$ & $\{t_{(r1, r2)}, t_{(r1, r3)}, t_{(r1, r4)},$\\
          	& $t_{(r2, r3)}, t_{(r2, r4)}, t_{(r3, r4)} \}$\\
			$(b,\uparrow)$ & $\{t_{(r1, r2)}, t_{(r1, r3)}, t_{(r1, r4)},$\\
          	& $t_{(r2, r3)}, t_{(r2, r4)}, t_{(r3, r4)} \}$\\   
      		$(c,\uparrow)$ & $\{t_{(r1, r4)}, t_{(r2, r3)}, t_{(r2, r4)},$\\
      		& $t_{(r3, r4)} \}$\\
      		$(d,\downarrow)$ & $\{t_{(r1, r2)}, t_{(r1, r4)}, t_{(r2, r4)},$\\
          	& $t_{(r3, r4)} \}$\\
      		\bottomrule
    	\end{tabular}}%
			\\
		\subfloat[]{%
		\begin{tabular}{|c|c c c c|}
			\hline
			$\Rsh$ & r1 & r2 & r3 & r4\\
			\hline
			r1 & 1 & $1/4$ & $1/3$ & $1/5$\\
			r2 & 1 & 1 & $1/4$ & $1/5$\\
			r3 & 1 & 1 & 1 & $1/5$\\
			r4 & 1 & 1 & 1 & 1\\
			\hline
			\end{tabular}}%
    	\label{tab4:trans_costmatrix}
	\end{table}
	
	We define a 2-Dimensional \textbf{cost matrix} whose size is: $\mathcal{C}_{i,j} = n\times n$ (where $n$ is number of tuples in the original numeric data set, such assuch as Table~\ref{tab4:sample4}) and initialize it to 1. Using the sorted and reduced transactional data set in Table~\ref{tab4:trans_costmatrix}a, we update elements of \textbf{cost matrix} that correspond to the inverse of occurrence count of \textbf{tids} as shown in Equation~\eqref{eqn4:c-matrix}. Table~\ref{tab4:trans_costmatrix}b illustrates the \textbf{cost matrix} of Table~\ref{tab4:trans_costmatrix}a. In this case we observe that \textbf{tids} with more occurrences have least costs; therefore, they are better candidates for parent nodes.
	
	\begin{equation}\label{eqn4:c-matrix}
		\mathcal{C}_{i, j} = \frac{1}{1 + \sum (ri, rj)_{count}}
	\end{equation}
	\begin{center}
		\begin{footnotesize}where $(ri, rj)_{count}$ is the occurrence count of tid pair $(ri, rj)$.\end{footnotesize}
	\end{center}
	
	We define an artificial 2-Dimensional pheromone matrix whose size is shown in Equation~\eqref{eqn4:p-matrix_paraminer}. This matrix contains knowledge about the pheromone proportions of every node of the data set.
	
	\begin{equation}\label{eqn4:p-matrix_paraminer}
		\Gamma_{i, j} = n \times n
	\end{equation}
	\begin{center}
		where n: number of tuples in numeric data set
	\end{center}
		
	At the beginning all artificial pheromones $p_{i, j}$ in matrix $\Gamma_{i, j}$ are initialized to 1, then they are updated as follows at time interval $(t, t+1)$:
	
	\begin{itemize}
		\item every generated node, is used to retrieve the corresponding attributes from the sorted reduced transactional data set. A set intersection of the \textbf{tids} of all these attributes provides an FP-Tree whose length is tested against the specified threshold;
		\item if the length of the FP-tree surpasses the specified threshold, the pheromones corresponding to the \textbf{tids} are incremented by 1: $p_{i,j} = p_{i,j}(t) + 1$;
		\item if the length of the FP-Tree falls below the specified threshold, the pheromones corresponding to the \textbf{tids} are evaporated by a factor: $(1 - \rho)\cdot p_{i, j}(t)$.
	\end{itemize}
	
	We propose a probabilistic rule that allows us to learn a parent node of the longest FP-Tree in a non-recursive manner. This rule is shown in Equation~\eqref{eqn4:acoparaminer_rule}.
	
	\begin{equation}\label{eqn4:acoparaminer_rule}
		\mathcal{P}_{i, j}(t) = \frac{p_{i, j} (1 / \mathcal{C}_{i, j})}{\sum _{k=1} ^{n} p_{i, j} ^{k} (t) (1 / \mathcal{C}_{i, j} ^{k})}
	\end{equation}
		
	The main steps for the ACO-ParaMiner algorithm are:
	
	\begin{enumerate}
		\item Encode a data set $\mathcal{D}$ into a transactional data set and reduce it by combining similar item sets and the removing infrequent ones. Use the reduced transactional data set $\mathcal{D}_{red}$ to construct a Cost Matrix $\mathcal{C}$.
		\item Generate nodes that have high probability of being parent nodes using the Cost Matrix and Pheromone Matrix $\Gamma_{i,j}$. 
		\item For each generated node, retrieve corresponding attributes from $\mathcal{D}_{red}$ and determine the intersection of their \texttt{tids}. If the length of the resulting intersection surpasses the minimum length threshold $\varsigma$, then update the pheromones.
		\item Repeat steps 2-3 until the algorithm starts to generate similar nodes.
	\end{enumerate}
	
	\subsubsection{Computational Complexity of ACO-ParaMiner}
	\label{sec4.3.3}
	We use big-O notation \cite{Cormen2009,Vaz2017} to analyze the computational complexity of ACO-ParaMiner algorithm. Firstly, ACO-ParaMiner constructs a \textbf{cost matrix} of all corresponding nodes. Secondly, ACO-ParaMiner learns the longest FP-Tree with the help of the \textbf{cost matrix}. We formulate the problem and show its complexity.
	
	\textbf{Problem formulation.} Given a dataset $\mathcal{D}$ with $m$ attributes and $n$ objects, an encoded transactional data set is built and used to construct a \textbf{cost matrix} of size $n\times n$. In order to build an encoded transactional data set (as shown in Table~\ref{tab4:transactional2}a) every tuple object is paired with all other tuple objects and the resulting gradual patterns recorded. Each tuple pairing results into an iteration that identifies gradual patterns; therefore, the computation complexity is equal to the total number of pairings ($O(\frac{n^{2} - n}{2})$). Next, using this data set \textbf{tids} are generated for every possible singleton gradual item and those whose length do not satisfy the user-defined threshold are discarded (as shown in Table~\ref{tab4:transactional2}b and Table~\ref{tab4:trans_costmatrix}a respectively). It should be remembered that for every attribute, 2 singleton gradual items are possible (for example, for attribute $a$ we have $(a, \uparrow)$ and $(a, \downarrow)$). Every singleton gradual item results into an iteration that finds \textbf{tids} from the previous data set and may be equated to a constant complexity of $C_{1}$; therefore, complexity is equal to the total number of the items ($O(2m\cdot C_{1})$). The computational complexity of building a \textbf{cost matrix} from resulting transactional data set is minuscule in comparison to the previous processes. On that account, the total complexity is $O(\frac{n^{2} - n}{2} + 2m\cdot C_{1})$.
		
	\textbf{Search space size.} According to \cite{Negrevergne2014}, the process of finding the longest FP-Tree using a transactional data set is recursive. It can be shown that a recursive function introduces a large search space which makes it computationally expensive. The aim of ACO-ParaMiner is to find the longest FP-Tree using a non-recursive function. As a result, if an ACO-ParaMiner algorithm iterates $x$ times it generates $x$ FP-Trees. Every FP-Tree that is generated, is applied on the reduced encoded transactional data set (as shown in Table~\ref{tab4:trans_costmatrix}a) in order to \textit{join} all singleton gradual items that share that FP-Tree to form a gradual pattern. Next, the \textbf{tid} length of the formed gradual pattern compared against a user-defined threshold. This process may be equated to a constant complexity of $C_{2}$. The overall complexity of ACO-ParaMiner becomes $O(\frac{n^{2} - n}{2} + 2m\cdot C_{1} + x\cdot C_{2})$.
	
	Notably, although complexities of $C_{1}$ and $C_{1}$ are almost constant for every iteration; in the long run, they produce a significant computational expense.

	\section{Experiments}
	\label{sec5:experiments}
	In this section, we present an experimental study of computational performance of our algorithms. We implemented BFS-based GRAANK (as described in \cite{Laurent2009}) and ACO-GRAANK; and DFS-based ParaMiner (as described in \cite{Negrevergne2014}) and ACO-ParaMiner algorithms in \texttt{Python} language. These ACO-based algorithms can also be described as algorithms that are built on top of a population-based optimization technique (that is ACO). As discussed in Section~\ref{sec3:related}, GA and PSO are alternative population-based optimization techniques which may also be applied in GP mining. As a result, we implement GA-GP and PSO-GP algorithms as alternative algorithms to the ACO-based algorithms and we perform experiments in order to compare their computational performances. In addition to these experiments, we also compare the performance of these population-based algorithms against their classical GP mining algorithm counterparts (that is GRAANK and ParaMiner). All experiments were conducted on a HPC (High Performance Computing) platform \textbf{Meso@LR}\footnote{\url{https://meso-lr.umontpellier.fr}}. We used one node made up of 14 cores and 128GB of RAM.
	
	\subsection{Source Code}
	\label{sec5.1:source_code}
	The \textbf{Python} (language) source code of the 6 algorithms is available at our GitHub repository: \begin{footnotesize}\url{https://github.com/owuordickson/ant-colony-gp.git}\end{footnotesize}. Since \cite{Negrevergne2014} does not provide a Python implementation of ParaMiner, we extend a LCM Python source code \begin{footnotesize}(\url{https://github.com/scikit-mine.git})\end{footnotesize} to implement our ParaMiner algorithm.

	\subsection{Data set Description}
	\label{sec5.2:dataset}

	\begin{table}[h!]
	\scriptsize
	\centering
	\caption{Experiment data sets.}
	\begin{tabular}{|c|c|c|c|}
		\hline
		Data set & $\#$tuples & $\#$attributes & Domain\\
		\hline \hline
		Breast Cancer (B\&C) & 116 & 10 & Medical\\
		Cargo 2000 (C2K) & 3942 & 98 & Transport\\
		Directio (Buoys) & 948000 & 21 & Coastline\\
		Power Consump. (UCI) & 2075259 & 9 & Electrical\\
		\hline
	\end{tabular}
	\label{tab:dataset}
	\end{table}
	
	Table~\ref{tab:dataset} shows the features of the data sets used in the experiments for evaluating the computational performance of the algorithms. All the data sets described in this table are numerical data sets. The `Breast Cancer' data set, obtained from UCI Machine Learning Repository (UCI-MLR) \cite{Patricio2018}, is composed of 10 quantitative predictors and binary variable indicating the presence or absence of breast cancer. The predictors are recordings of anthropometric data gathered from the routine blood analysis of 116 participants (of whom 64 have breast cancer and 52 are healthy).
	
	The `Cargo 2000' data set,  obtained from UCI-MLR \cite{Metzger2015}, describes 98 tracking and tracing events that span 5 months of transport logistics execution. The `Power Consumption' data set, obtained from UCI-MLR \cite{Dua2019}, describes the electric power consumption in one household (located in Sceaux, France) in terms of active power, voltage and global intensity with a one-minute sampling rate between 2006 and 2010.
	
	The `Directio' data set is one of 4 data sets obtained from OREMES’s data portal\footnote{\url{https://data.oreme.org}} that recorded swell sensor signals of 4 buoys near the coast of the Languedoc-Roussillon region in France between 2012 and 2019 \cite{Bouchette2019}. 

	\subsection{Experiment Results}
	\label{sec5.3:results}
	In this section, we present our experiment results which reveal the computational behavior of GRAANK, Para-Miner, ACO-GRAANK, ACO-ParaMiner, GA-GP, \\PSO-GP algorithms when applied on the 4 data sets described in Section~\ref{sec5.2:dataset}. The results are a mean different test runs of each algorithm on each data set at different parameter settings. All these results are available at:\begin{footnotesize} \url{https://github.com/owuordickson/meso-hpc-lr/tree/master/results/swarm}\end{footnotesize}.

	\subsubsection{Experiment 1: Population-based Algorithms}
	\label{sec:5.3.1:swarm_intel}
\begin{figure}[h!]
		\centering
		\small
		\begin{tikzpicture}
			\begin{axis}[
			height=4.5cm, width=7.5cm,
  			grid=both,
  			grid style={line width=.1pt, draw=gray!20},
    		major grid style={line width=.2pt, draw=gray!50},
  			axis lines=middle,
  			minor tick num=5,
    		ymin=0, ymax=70,
    		xmin=0.48, xmax=0.95,
    		xtick={0.5, 0.6, 0.7, 0.8, 0.9},
   			xlabel=Minimum support threshold $(\sigma)$, 
   			ylabel=GP Count (mean),,
  			xlabel style={at={(axis description cs:0.5,-0.1)},anchor=north},
  			ylabel style={at={(axis description cs:-0.15,0.5)},rotate=90, anchor=south},
   	 	 	legend style={at={(0.8,0.8)}, nodes={scale=0.5}}
 			]

			\addplot[smooth, mark=x, mark size=2pt, color=red] table[x=Support, y=ACG]{bc_pattern.dat};
  	  		\addlegendentry{ACG}
			
			\addplot[mark=square, mark size=2pt, color=brown] table[x=Support, y=GRA]{bc_pattern.dat};
  	  		\addlegendentry{GRA}
  	  		
  	  		\addplot[mark=o, mark size=2pt, color=blue] table[x=Support, y=GAG]{bc_pattern.dat};
  	  		\addlegendentry{GAG}
  	  		
  	  		\addplot[mark=triangle, mark size=2pt, color=black] table[x=Support, y=PSO]{bc_pattern.dat};
  	  		\addlegendentry{PSO}
  	  		
  	  		\addplot[mark=*, mark size=2pt, color=purple] table[x=Support, y=PRM]{bc_pattern.dat};
  	  		\addlegendentry{PRM}
  	  		
  	  		\addplot[mark=+, mark size=2pt, color=cyan] table[x=Support, y=ACP]{bc_pattern.dat};
  	  		\addlegendentry{ACP}
  	  		  	  		
  			\end{axis}
  		\end{tikzpicture}
  		
    	\caption{\textbf{B\&C data set:} plot of GP count against minimum support.}
  		\label{fig5:swarmgp_exp}
  	\end{figure}
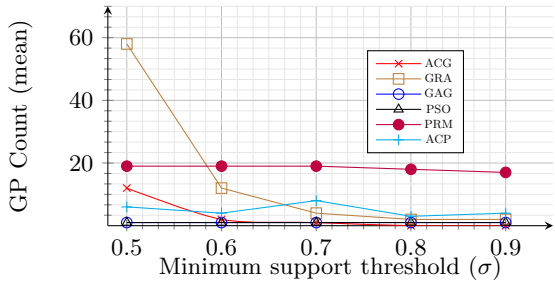

	In this experiment we tune the respective parameters of ACO-GRAANK, GA-GP, and PSO-GP algorithms and apply each to GP mining. We apply the 3 algorithms on the 4 data sets described in Section~\ref{sec5.2:dataset}, we set the maximum iteration count to 100, minimum support threshold $(\sigma)$ to 0.5 and we observe their behaviors as they mine for GPs. Each test run is repeated several times in order to confirm each behavior. We compute the mean count of extracted GPs by each algorithm and plot the results in Figure~\ref{fig5:swarmgp_exp}. More detailed results are shown in Tables~\ref{tab5:exp2_bc}, \ref{tab5:exp2_buoys}, \ref{tab5:exp2_c2k} and \ref{tab5:exp2_uci}.
	
	It is important to note that for the case of ACO-GRAANK, GA-GP, and PSO-GP, the term `best cost' or `best position' refers to a value that is computed after every iteration using the respective fitness function which aims to find GPs with high frequency support. In other words, the higher the frequency support of a GP, the lower its `cost' or `position'.
	
	In the case of ACO-GRAANK algorithm, we varied the evaporation factor value between 0.1 to 0.9. It is important to recall that evaporation factor determines how fast unused pheromone trails disappear (see Section~\ref{sec4.1:aco}). The lower the evaporation factor value, the faster the unused pheromone trails disappear. In this experiment, ACO-GRAANK algorithm fetches almost the same number of GPs regardless of the evaporation factor value.
	
	In the case of GA-GP, we varied the `Proportion of Children' (PC) factor value between 0.1 to 0.9. PC value determines the population size of the offsprings/ children $(nc)$ that are produced at the crossover stage ($nc = PC\cdot Npop$) \cite{Mirjalili2019}. As long as the PC factor value was above 0.1, GA-GP algorithm finds GP with the highest frequency support in a few iterations. This may imply that increasing PC value increases the chances of producing higher quality offsprings whose fitness correspond to finding GPs with high frequency supports.

	In the case of PSO-GP, we varied coefficients C1 and C2 values between 0.1 to 0.9. C2 coefficient value gives the importance of the global best position, while C1 coefficient value gives the importance of the personal best position \cite{Shruti2012}. Regardless of the values of C1 and C2 coefficients, PSO-GP algorithm finds almost the same number of GPs which have highest frequency support within a few iterations.

	Given these results, we observe that ACO-GRAANK algorithm is best suited for finding numerous frequent GPs, while GA-GP and PSO-GP are best suited for finding only GPs with the highest frequency support.

	\subsubsection{Experiment 2: Memory Utilization}
	\label{sec5.3.2:memory}
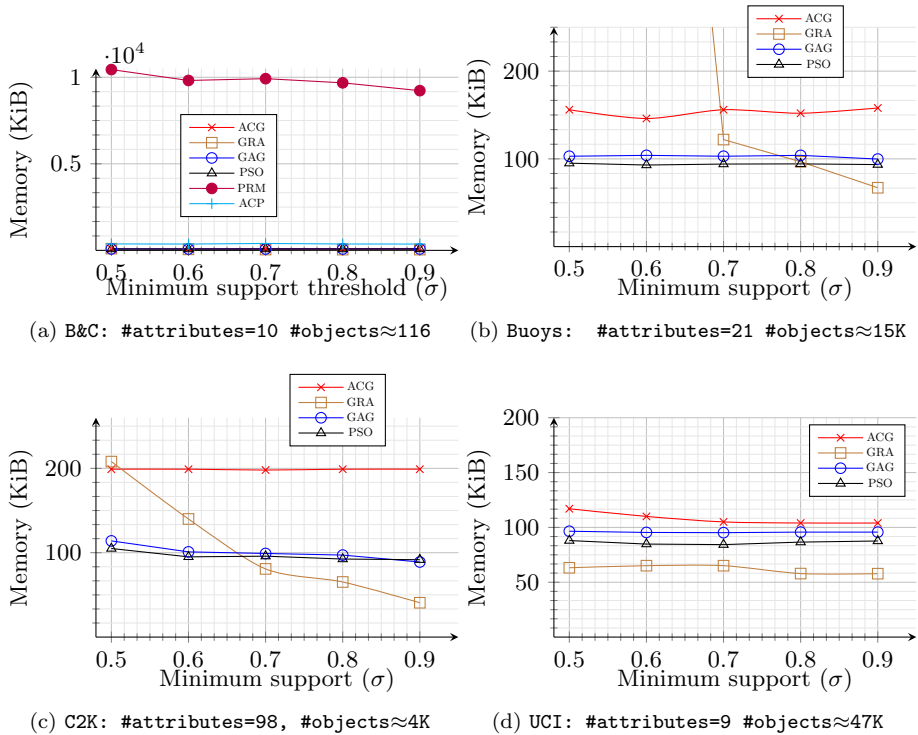
\begin{figure}[h!]
		\centering
		\small
		\subfloat[\texttt{B\&C: \#attributes=10 \#objects$\approx$116}]{%
		\begin{tikzpicture}
  			\begin{axis}[
  				height=4cm, width=6.4cm,
  				grid=both,
  				grid style={line width=.1pt, draw=gray!20},
    			major grid style={line width=.2pt, draw=gray!50},
  				axis lines=middle,
  				minor tick num=5,
    			ymin=0, ymax=10500,
    			xmin=0.48, xmax=0.95,
    			xtick={0.5, 0.6, 0.7, 0.8, 0.9},
   				xlabel=Minimum support threshold $(\sigma)$, 
   				ylabel=Memory (KiB),
  				xlabel style={at={(axis description cs:0.5,-0.1)},anchor=north},
  				ylabel style={at={(axis description cs:-0.15,0.5)},rotate=90, anchor=south},
   	 	 		legend style={at={(0.5,0.75)}, nodes={scale=0.5}}
 				]

			\addplot[smooth, mark=x, mark size=2pt, color=red] table[x=Support, y=ACG]{bc_memory.dat};
  	  		\addlegendentry{ACG}
			
			\addplot[mark=square, mark size=2pt, color=brown] table[x=Support, y=GRA]{bc_memory.dat};
  	  		\addlegendentry{GRA}
  	  		
  	  		\addplot[mark=o, mark size=2pt, color=blue] table[x=Support, y=GAG]{bc_memory.dat};
  	  		\addlegendentry{GAG}
  	  		
  	  		\addplot[mark=triangle, mark size=2pt, color=black] table[x=Support, y=PSO]{bc_memory.dat};
  	  		\addlegendentry{PSO}
  	  		
  	  		\addplot[mark=*, mark size=2pt, color=purple] table[x=Support, y=PRM]{bc_memory.dat};
  	  		\addlegendentry{PRM}
  	  		
  	  		\addplot[mark=+, mark size=2pt, color=cyan] table[x=Support, y=ACP]{bc_memory.dat};
  	  		\addlegendentry{ACP}
  			\end{axis}
  		\end{tikzpicture}}%
		\subfloat[\texttt{Buoys: \#attributes=21 \#objects$\approx$15K}]{%
		\begin{tikzpicture}
  			\begin{axis}[
  				height=4.5cm, width=6.4cm,
  				grid=both,
  				grid style={line width=.1pt, draw=gray!20},
    			major grid style={line width=.2pt, draw=gray!50},
  				axis lines=middle,
  				minor tick num=5,
    			ymin=0, ymax=250,
    			xmin=0.48, xmax=0.95,
    			xtick={0.5, 0.6, 0.7, 0.8, 0.9},
   				xlabel=Minimum support $(\sigma)$, 
   				ylabel=Memory (KiB),
  				xlabel style={at={(axis description cs:0.5,-0.1)},anchor=north},
  				ylabel style={at={(axis description cs:-0.15,0.5)},rotate=90, anchor=south},
   	 	 		legend style={at={(0.8,1.1)}, nodes={scale=0.5}}
 				]

			\addplot[smooth, mark=x, mark size=2pt, color=red] table[x=Support, y=ACG]{buoys_memory.dat};
  	  		\addlegendentry{ACG}
			
			\addplot[mark=square, mark size=2pt, color=brown] table[x=Support, y=GRA]{buoys_memory.dat};
  	  		\addlegendentry{GRA}
  	  		
  	  		\addplot[mark=o, mark size=2pt, color=blue] table[x=Support, y=GAG]{buoys_memory.dat};
  	  		\addlegendentry{GAG}
  	  		
  	  		\addplot[mark=triangle, mark size=2pt, color=black] table[x=Support, y=PSO]{buoys_memory.dat};
  	  		\addlegendentry{PSO}
  			\end{axis}
  		\end{tikzpicture}}%

		\subfloat[\texttt{C2K: \#attributes=98, \#objects$\approx$4K}]{%
		\begin{tikzpicture}
  			\begin{axis}[
  				height=4.5cm, width=6.4cm,
  				grid=both,
  				grid style={line width=.1pt, draw=gray!20},
    			major grid style={line width=.2pt, draw=gray!50},
  				axis lines=middle,
  				minor tick num=5,
    			ymin=0, ymax=260,
    			xmin=0.48, xmax=0.95,
    			xtick={0.5, 0.6, 0.7, 0.8, 0.9},
   				xlabel=Minimum support $(\sigma)$, 
   				ylabel=Memory (KiB),
  				xlabel style={at={(axis description cs:0.5,-0.1)},anchor=north},
  				ylabel style={at={(axis description cs:-0.15,0.5)},rotate=90, anchor=south},
   	 	 		legend style={at={(0.8,1.2)}, nodes={scale=0.5}},
 				]

			\addplot[smooth, mark=x, mark size=2pt, color=red] table[x=Support, y=ACG]{c2k_memory.dat};
  	  		\addlegendentry{ACG}
			
			\addplot[smooth, mark=square, mark size=2pt, color=brown] table[x=Support, y=GRA]{c2k_memory.dat};
  	  		\addlegendentry{GRA}
  	  		
  	  		\addplot[mark=o, mark size=2pt, color=blue] table[x=Support, y=GAG]{c2k_memory.dat};
  	  		\addlegendentry{GAG}
  	  		
  	  		\addplot[mark=triangle, mark size=2pt, color=black] table[x=Support, y=PSO]{c2k_memory.dat};
  	  		\addlegendentry{PSO}
			\end{axis}
  		\end{tikzpicture}}%
  		\subfloat[\texttt{UCI: \#attributes=9 \#objects$\approx$47K}]{%
		\begin{tikzpicture}
  			\begin{axis}[
  				height=4.5cm, width=6.4cm,
  				grid=both,
  				grid style={line width=.1pt, draw=gray!20},
    			major grid style={line width=.2pt, draw=gray!50},
  				axis lines=middle,
  				minor tick num=5,
    			ymin=0, ymax=200,
    			xmin=0.48, xmax=0.95,
    			xtick={0.5, 0.6, 0.7, 0.8, 0.9},
   				xlabel=Minimum support $(\sigma)$, 
   				ylabel=Memory (KiB),
  				xlabel style={at={(axis description cs:0.5,-0.1)},anchor=north},
  				ylabel style={at={(axis description cs:-0.15,0.5)},rotate=90, anchor=south},
 	 			legend style={nodes={scale=0.5}},
 	 			legend pos=north east
 				]

			\addplot[smooth, mark=x, mark size=2pt, color=red] table[x=Support, y=ACG]{uci_memory.dat};
  	  		\addlegendentry{ACG}
  	  					
			\addplot[smooth, mark=square, mark size=2pt, color=brown] table[x=Support, y=GRA]{uci_memory.dat};
  	  		\addlegendentry{GRA}
  	  		
  	  		\addplot[mark=o, mark size=2pt, color=blue] table[x=Support, y=GAG]{uci_memory.dat};
  	  		\addlegendentry{GAG}
  	  		
  	  		\addplot[mark=triangle, mark size=2pt, color=black] table[x=Support, y=PSO]{uci_memory.dat};
  	  		\addlegendentry{PSO}
  			\end{axis}
  		\end{tikzpicture}}%
  		
    	\caption{Plot of memory usage against minimum support.}
  		\label{fig5:exp1_memory}
  	\end{figure}

	This experiment identifies the algorithms with better memory utilization while extracting GPs. For this experiment, we apply ACO-GRAANK (ACG), ACO-Para-Miner (ACP), GA-GP (GAG), GRAANK (GRA), Para-Miner and PSO-GP (PSO) algorithms on B\&C, Buoys, C2K and UCI data sets. We perform multiple test runs of each algorithm on every data set as we vary minimum support threshold $(\sigma)$ from 0.5 to 0.9, we compute the mean of allocated memory-size values and we plot the graphs shown in Figure~\ref{fig5:exp1_memory}. More detailed results are shown in Tables~\ref{tab5:exp2_bc}, \ref{tab5:exp2_buoys}, \ref{tab5:exp2_c2k} and \ref{tab5:exp2_uci}.
	
  	 In Figure~\ref{fig5:exp1_memory}a, we observe that ACP algorithm requires the highest memory allocation. It is important to mention that ACP and PRM algorithms yield an \textit{`Out of Memory Error'} and no run-time when applied on Buoys, C2K and UCI data sets. GRAANK yields the same error when applied on Buoys data set with at least $15,0000$ tuples (Figure~\ref{fig5:exp1_memory}b) and UCI data set with more than $47,000$ tuples. ACG, GAG and PSO do not yield an \textit{Out of Memory Error} in any of these experiments as shown in Figures~\ref{fig5:exp1_memory}a, \ref{fig5:exp1_memory}b, \ref{fig5:exp1_memory}c, \ref{fig5:exp1_memory}d.

	\subsubsection{Experiment 3: Computational Performance}
	\label{sec5.3.3:computational}
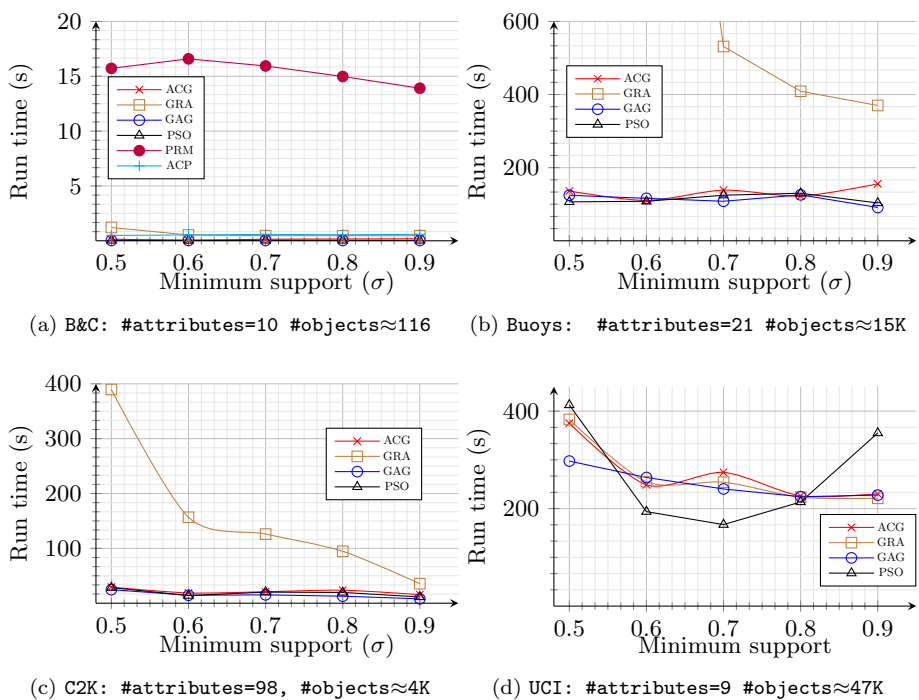
\begin{figure}[h!]
		\centering
		\small
		\subfloat[\texttt{B\&C: \#attributes=10 \#objects$\approx$116}]{%
		\begin{tikzpicture}
  			\begin{axis}[
  				height=4.5cm, width=6.4cm,
  				grid=both,
  				grid style={line width=.1pt, draw=gray!20},
    			major grid style={line width=.2pt, draw=gray!50},
  				axis lines=middle,
  				minor tick num=5,
    			ymin=0, ymax=20,
    			xmin=0.48, xmax=0.95,
    			xtick={0.5, 0.6, 0.7, 0.8, 0.9},
   				xlabel=Minimum support $(\sigma)$, 
   				ylabel=Run time (s),
  				xlabel style={at={(axis description cs:0.5,-0.1)},anchor=north},
  				ylabel style={at={(axis description cs:-0.15,0.5)},rotate=90, anchor=south},
   	 	 		legend style={at={(0.3,0.75)}, nodes={scale=0.5}}
 				]

			\addplot[smooth, mark=x, mark size=2pt, color=red] table[x=Support, y=ACG]{bc_runtime.dat};
  	  		\addlegendentry{ACG}
			
			\addplot[mark=square, mark size=2pt, color=brown] table[x=Support, y=GRA]{bc_runtime.dat};
  	  		\addlegendentry{GRA}
  	  		
  	  		\addplot[mark=o, mark size=2pt, color=blue] table[x=Support, y=GAG]{bc_runtime.dat};
  	  		\addlegendentry{GAG}
  	  		
  	  		\addplot[mark=triangle, mark size=2pt, color=black] table[x=Support, y=PSO]{bc_runtime.dat};
  	  		\addlegendentry{PSO}
  	  		
  	  		\addplot[mark=*, mark size=2pt, color=purple] table[x=Support, y=PRM]{bc_runtime.dat};
  	  		\addlegendentry{PRM}
  	  		
  	  		\addplot[mark=+, mark size=2pt, color=cyan] table[x=Support, y=ACP]{bc_runtime.dat};
  	  		\addlegendentry{ACP}
  			\end{axis}
  		\end{tikzpicture}}%
		\subfloat[\texttt{Buoys: \#attributes=21 \#objects$\approx$15K}]{%
		\begin{tikzpicture}
  			\begin{axis}[
  				height=4.5cm, width=6.4cm,
  				grid=both,
  				grid style={line width=.1pt, draw=gray!20},
    			major grid style={line width=.2pt, draw=gray!50},
  				axis lines=middle,
  				minor tick num=5,
    			ymin=0, ymax=600,
    			xmin=0.48, xmax=0.95,
    			xtick={0.5, 0.6, 0.7, 0.8, 0.9},
   				xlabel=Minimum support $(\sigma)$, 
   				ylabel=Run time (s),
  				xlabel style={at={(axis description cs:0.5,-0.1)},anchor=north},
  				ylabel style={at={(axis description cs:-0.15,0.5)},rotate=90, anchor=south},
   	 	 		legend style={at={(0.3,0.8)}, nodes={scale=0.5}}
 				]

			\addplot[smooth, mark=x, mark size=2pt, color=red] table[x=Support, y=ACG]{buoys_runtime.dat};
  	  		\addlegendentry{ACG}
			
			\addplot[mark=square, mark size=2pt, color=brown] table[x=Support, y=GRA]{buoys_runtime.dat};
  	  		\addlegendentry{GRA}
  	  		
  	  		\addplot[mark=o, mark size=2pt, color=blue] table[x=Support, y=GAG]{buoys_runtime.dat};
  	  		\addlegendentry{GAG}
  	  		
  	  		\addplot[mark=triangle, mark size=2pt, color=black] table[x=Support, y=PSO]{buoys_runtime.dat};
  	  		\addlegendentry{PSO}
  			\end{axis}
  		\end{tikzpicture}}%

		\subfloat[\texttt{C2K: \#attributes=98, \#objects$\approx$4K}]{%
		\begin{tikzpicture}
  			\begin{axis}[
  				height=4.5cm, width=6.4cm,
  				grid=both,
  				grid style={line width=.1pt, draw=gray!20},
    			major grid style={line width=.2pt, draw=gray!50},
  				axis lines=middle,
  				minor tick num=5,
    			ymin=0, ymax=400,
    			xmin=0.48, xmax=0.95,
    			xtick={0.5, 0.6, 0.7, 0.8, 0.9},
   				xlabel=Minimum support $(\sigma)$, 
   				ylabel=Run time (s),
  				xlabel style={at={(axis description cs:0.5,-0.1)},anchor=north},
  				ylabel style={at={(axis description cs:-0.15,0.5)},rotate=90, anchor=south},
   	 	 		legend style={at={(0.9,0.8)}, nodes={scale=0.5}}
 				]

			\addplot[smooth, mark=x, mark size=2pt, color=red] table[x=Support, y=ACG]{c2k_runtime.dat};
  	  		\addlegendentry{ACG}
			
			\addplot[smooth, mark=square, mark size=2pt, color=brown] table[x=Support, y=GRA]{c2k_runtime.dat};
  	  		\addlegendentry{GRA}
  	  		
  	  		\addplot[mark=o, mark size=2pt, color=blue] table[x=Support, y=GAG]{c2k_runtime.dat};
  	  		\addlegendentry{GAG}
  	  		
  	  		\addplot[mark=triangle, mark size=2pt, color=black] table[x=Support, y=PSO]{c2k_runtime.dat};
  	  		\addlegendentry{PSO}
			\end{axis}
  		\end{tikzpicture}}%
  		\subfloat[\texttt{UCI: \#attributes=9 \#objects$\approx$47K}]{%
		\begin{tikzpicture}
  			\begin{axis}[
  				height=4.5cm, width=6.4cm,
  				grid=both,
  				grid style={line width=.1pt, draw=gray!20},
    			major grid style={line width=.2pt, draw=gray!50},
  				axis lines=middle,
  				minor tick num=5,
    			ymin=0, ymax=450,
    			xmin=0.48, xmax=0.95,
    			xtick={0.5, 0.6, 0.7, 0.8, 0.9},
   				xlabel=Minimum support, 
   				ylabel=Run time (s),
  				xlabel style={at={(axis description cs:0.5,-0.1)},anchor=north},
  				ylabel style={at={(axis description cs:-0.15,0.5)},rotate=90, anchor=south},
 	   	 	 	legend style={at={(1,0.42)}, nodes={scale=0.5}},
 				]

			\addplot[smooth, mark=x, mark size=2pt, color=red] table[x=Support, y=ACG]{uci_runtime.dat};
  	  		\addlegendentry{ACG}
  	  					
			\addplot[smooth, mark=square, mark size=2pt, color=brown] table[x=Support, y=GRA]{uci_runtime.dat};
  	  		\addlegendentry{GRA}
  	  		
  	  		\addplot[mark=o, mark size=2pt, color=blue] table[x=Support, y=GAG]{uci_runtime.dat};
  	  		\addlegendentry{GAG}
  	  		
  	  		\addplot[mark=triangle, mark size=2pt, color=black] table[x=Support, y=PSO]{uci_runtime.dat};
  	  		\addlegendentry{PSO}
  			\end{axis}
  		\end{tikzpicture}}%
    	\caption{Plot of run-time against minimum support.}
  		\label{fig5:exp1_runtime}
  	\end{figure}

	Similar to Section~\ref{sec5.3.2:memory},  this experiment compares the computational performance of the 6 algorithms (in terms of run-time) when applied on the 4 data sets. Using the results, we plot graphs in Figure~\ref{fig5:exp1_runtime} and more detailed results are shown in Tables~\ref{tab5:exp2_bc}, \ref{tab5:exp2_buoys}, \ref{tab5:exp2_c2k}, \ref{tab5:exp2_uci}.
	
	In Figure~\ref{fig5:exp1_runtime}a, we observe that PRM algorithm slowest run-time in comparison to ACG, ACP, GAG and GRA algorithms. In Figure~\ref{fig5:exp1_runtime}b, GRA yields an \textit{`Out of Memory Error'} when $\sigma \leq 0.5$ and it has the slowest run-time when $\sigma \geq 0.7$. Similarly In Figure~\ref{fig5:exp1_runtime}c, GRA has the slowest run-time that reduces as $\sigma$ increases. In Figure~\ref{fig5:exp1_runtime}d, ACG, GRA, GAG and PSO algorithms have almost the same run-times. It should be noted that ACP and PRM algortithms yield an \textit{`Out of Memory Error'} when applied to Buoys, C2K and UCI data sets.
	
	\newpage
	
	\begin{sidewaystable}
  		\centering
    	\caption{\textbf{B\&C data set:} computational comparison of ACO-GRAANK (ACG), ACO-ParaMiner (ACP), GA-GP (GAG), GRAANK (GRA), ParaMiner (PRM) and PSO-GP (PSO) at different support (sup) values.}
    	\begin{tabular}{c c | c c c c : c c c c : c c c c}
      	\multirow{2}{*}{\textbf{Sup}} & \multirow{2}{*}{\textbf{Alg}} &
      	\multicolumn{4}{c}{\textbf{Run-time (sec)}} &
      	\multicolumn{4}{c}{\textbf{No. of Gradual Patterns}} &
      	\multicolumn{4}{c}{\textbf{Memory (KiB)}}\\
		& & Std.Dev & Best & Mean & Worse & Std.Dev & Fewest & Mean & Most & Std.Dev & Min & Mean & Max\\
      	\toprule
      	0.5 & ACG & 0.047 & 0.149 & 0.204 & 0.231 & 2.000 & 12.0 & 14.000 & 16.0 & 3.786 & 103.0 & 107.333 & 110.0\\
			& ACP & 0.051 & 0.491 & 0.546 & 0.591 & 7.638 & 6.0 & 14.333 & 21.0 & 21.455 & 373.0 & 397.667 & 412.0\\
			& GAG & \textbf{0.006} & \textbf{0.034} & \textbf{0.040} & \textbf{0.046} & 0.577 & 1.0 & 1.333 & 2.0 & 0.981 & 84.9 & 85.467 & 86.6\\
			& GRA & 0.006 & 1.212 & 1.219 & 1.223 & \textbf{0.000} & \textbf{58.0} & \textbf{58.000} & \textbf{58.0} & 0.462 & 98.6 & 99.133 & 99.4\\
			& PRM & 0.081 & 15.720 & 15.807 & 15.880 & 0.000 & 19.0 & 19.000 & 19.0 & 0.000 & 1044.0 & 1044.000 & 1044.0\\
			& PSO & 0.005 & 0.036 & 0.041 & 0.044 & 1.000 & 1.0 & 2.000 & 3.0 & \textbf{1.044} & \textbf{75.6} & \textbf{76.800} & \textbf{77.5}\\
		\hline
		0.6 & ACG & 0.080 & 0.055 & 0.143 & 0.212 & 2.082 & 2.0 & 4.333 & 6.0 & 8.327 & 93.0 & 102.333 & 109.0\\
			& ACP & 0.047 & 0.530 & 0.584 & 0.613 & 5.292 & 4.0 & 10.000 & 14.0 & 19.925 & 365.0 & 388.000 & 400.0\\
			& GAG & 0.003 & 0.031 & 0.035 & 0.037 & 0.000 & 1.0 & 1.000 & 1.0 & 2.307 & 80.5 & 83.133 & 84.8\\
			& GRA & 0.047 & 0.530 & 0.566 & 0.619 & 0.000 & 12.0 & 12.000 & 12.0 & \textbf{0.289} & \textbf{59.6} & \textbf{59.933} & \textbf{60.1}\\
			& PRM & 0.431 & 16.590 & 16.953 & 17.430 & \textbf{0.000} & \textbf{19.0} & \textbf{19.000} & \textbf{19.0} & 0.577 & 9814.0 & 9814.667 & 9815.0\\
			& PSO & \textbf{0.001} & \textbf{0.032} & \textbf{0.032} & \textbf{0.033} & 0.577 & 1.0 & 1.333 & 2.0 & 0.666 & 74.4 & 75.167 & 75.6\\
		\hline
		0.7 & ACG & 0.046 & 0.140 & 0.193 & 0.220 & 0.577 & 1.0 & 1.667 & 2.0 & 3.512 & 105.0 & 108.667 & 112.0\\
			& ACP & 0.050 & 0.586 & 0.640 & 0.684 & 6.506 & 8.0 & 14.667 & 21.0 & 16.503 & 399.0 & 417.333 & 431.0\\
			& GAG & \textbf{0.002} & \textbf{0.032} & \textbf{0.034} & \textbf{0.036} & 0.000 & 1.0 & 1.000 & 1.0 & 1.997 & 81.5 & 82.800 & 85.1\\
			& GRA & 0.007 & 0.469 & 0.475 & 0.483 & 0.000 & 4.0 & 4.000 & 4.0 & \textbf{0.058} & \textbf{47.4} & \textbf{47.467} & \textbf{47.5}\\
			& PRM & 0.285 & 15.940 & 16.227 & 16.510 & \textbf{0.000} & \textbf{19.0} & \textbf{19.000} & \textbf{19.0} & 0.577 & 9919.0 & 9919.333 & 9920.0\\
			& PSO & 0.003 & 0.035 & 0.037 & 0.040 & 0.000 & 1.0 & 1.000 & 1.0 & 0.551 & 75.5 & 76.033 & 76.6\\
		\hline
		0.8 & ACG & 0.022 & 0.165 & 0.190 & 0.208 & 0.577 & 0.0 & 0.667 & 1.0 & 2.517 & 106.0 & 108.333 & 111.0\\
			& ACP & 0.133 & 0.548 & 0.684 & 0.813 & 11.060 & 3.0 & 13.333 & 25.0 & 46.058 & 368.0 & 418.667 & 458.0\\
			& GAG & \textbf{0.002} & \textbf{0.031} & \textbf{0.033} & \textbf{0.034} & 0.000 & 1.0 & 1.000 & 1.0 & 0.265 & 80.7 & 80.900 & 81.2\\
			& GRA & 0.029 & 0.464 & 0.486 & 0.519 & 0.000 & 2.0 & 2.000 & 2.0 & \textbf{0.000} & \textbf{43.8} & \textbf{43.800} & \textbf{43.8}\\
			& PRM & 0.825 & 14.980 & 15.700 & 16.600 & \textbf{0.000} & \textbf{18.0} & \textbf{18.000} & \textbf{18.0} & 15.588 & 9678.0 & 9696.000 & 9705.0\\
			& PSO & 0.005 & 0.035 & 0.040 & 0.044 & 0.000 & 1.0 & 1.000 & 1.0 & 4.706 & 75.5 & 80.933 & 83.7\\
		\hline
		0.9 & ACG & 0.016 & 0.185 & 0.200 & 0.216 & 0.577 & 0.0 & 0.667 & 1.0 & 1.528 & 108.0 & 109.333 & 111.0\\
			& ACP & 0.164 & 0.589 & 0.779 & 0.882 & 10.440 & 4.0 & 16.000 & 23.0 & 35.360 & 365.0 & 405.333 & 431.0\\
			& GAG & \textbf{0.002} & \textbf{0.031} & \textbf{0.033} & \textbf{0.034} & 0.000 & 1.0 & 1.000 & 1.0 & 0.153 & 81.2 & 81.333 & 81.5\\
			& GRA & 0.041 & 0.461 & 0.488 & 0.536 & 0.000 & 2.0 & 2.000 & 2.0 & \textbf{0.000} & \textbf{43.8} & \textbf{43.800} & \textbf{43.8}\\
			& PRM & 0.784 & 13.910 & 14.463 & 15.360 & \textbf{0.000} & \textbf{17.0} & \textbf{17.000} & \textbf{17.0} & 0.000 & 9226.0 & 9226.000 & 9226.0\\
			& PSO & 0.004 & 0.031 & 0.036 & 0.039 & 0.000 & 1.0 & 1.000 & 1.0 & 0.929 & 79.1 & 80.133 & 80.9\\
      	\bottomrule
    	\end{tabular}
    	\label{tab5:exp2_bc}
	\end{sidewaystable}

	\begin{sidewaystable}
  		\centering
    	\caption{\textbf{Buoys data set:} computational comparison of ACG, GAG, GRA and PSO at different support (sup) values.}
    	\begin{tabular}{c c | c c c c : c c c c : c c c c}
      	\multirow{2}{*}{\textbf{Sup}} & \multirow{2}{*}{\textbf{Alg}} &
      	\multicolumn{4}{c}{\textbf{Run-time (sec)}} &
      	\multicolumn{4}{c}{\textbf{No. of Gradual Patterns}} &
      	\multicolumn{4}{c}{\textbf{Memory (KiB)}}\\
		& & Std.Dev & Best & Mean & Worse & Std.Dev & Fewest & Mean & Most & Std.Dev & Min & Mean & Max\\
      	\toprule
      	0.5 & ACG & 26.737 & 136.200 & 165.933 & 188.000 & \textbf{6.658} & \textbf{24.0} & \textbf{28.333} & \textbf{36.0} & 8.083 & 156.0 & 163.333 & 172.0\\
			& GAG & 16.626 & 124.500 & 140.133 & 157.600 & 0.577 & 1.0 & 1.333 & 2.0 & 2.887 & 103.0 & 104.667 & 108.0\\
			& PSO & \textbf{6.034} & \textbf{106.400} & \textbf{113.367} & \textbf{116.900} & 0.577 & 1.0 & 1.667 & 2.0 & \textbf{0.306} & \textbf{95.1} & \textbf{95.367} & \textbf{95.7}\\
		\hline
		0.6 & ACG & 23.426 & 110.200 & 130.633 & 156.200 & \textbf{2.646} & \textbf{16.0} & \textbf{18.000} & \textbf{21.0} & 9.074 & 146.0 & 156.333 & 163.0\\
			& GAG & \textbf{18.251} & \textbf{116.100} & \textbf{129.267} & \textbf{150.100} & 0.000 & 1.0 & 1.000 & 1.0 & 0.577 & 104.0 & 104.333 & 105.0\\
			& PSO & 42.289 & 108.000 & 139.033 & 187.200 & 0.577 & 1.0 & 1.667 & 2.0 & \textbf{0.819} & \textbf{93.0} & \textbf{93.900} & \textbf{94.6}\\
		\hline
		0.7 & ACG & 20.676 & 138.500 & 153.900 & 177.400 & 2.517 & 7.0 & 9.333 & 12.0 & 3.786 & 156.0 & 158.667 & 163.0\\
			& GAG & \textbf{9.582} & \textbf{108.100} & \textbf{118.933} & \textbf{126.300} & 0.000 & 1.0 & 1.000 & 1.0 & 0.577 & 103.0 & 103.333 & 104.0\\
			& GRA & 46.634 & 531.400 & 567.567 & 620.200 & \textbf{0.000} & \textbf{40.0} & \textbf{40.000} & \textbf{40.0} & 0.000 & 122.0 & 122.000 & 122.0\\
			& PSO & 22.652 & 124.700 & 144.300 & 169.100 & 0.577 & 1.0 & 1.667 & 2.0 & \textbf{0.866} & \textbf{94.1} & \textbf{95.100} & \textbf{95.6}\\
		\hline
		0.8 & ACG & 39.576 & 122.300 & 147.700 & 193.300 & 2.000 & 4.0 & 6.000 & 8.0 & 11.590 & 152.0 & 162.667 & 175.0\\
			& GAG & \textbf{5.575} & \textbf{125.600} & \textbf{129.767} & \textbf{136.100} & 0.000 & 1.0 & 1.000 & 1.0 & 0.000 & 104.0 & 104.000 & 104.0\\
			& GRA & 56.831 & 409.000 & 444.133 & 509.700 & \textbf{0.000} & \textbf{26.0} & \textbf{26.000} & \textbf{26.0} & 0.000 & 96.7 & 96.700 & 96.7\\
			& PSO & 32.057 & 130.000 & 160.900 & 194.000 & 0.577 & 1.0 & 1.667 & 2.0 & \textbf{0.907} & \textbf{94.3} & \textbf{95.267} & \textbf{96.1}\\
		\hline
		0.9 & ACG & 28.319 & 155.800 & 172.200 & 204.900 & 0.577 & 1.0 & 1.667 & 2.0 & 9.074 & 158.0 & 164.667 & 175.0\\
			& GAG & \textbf{10.334} & \textbf{91.050} & \textbf{99.980} & \textbf{111.300} & 0.577 & 1.0 & 1.333 & 2.0 & 1.793 & 99.8 & 100.933 & 103.0\\
			& GRA & 46.074 & 370.100 & 397.100 & 450.300 & \textbf{0.000} & \textbf{2.0} & \textbf{2.000} & \textbf{2.0} & \textbf{0.000} & \textbf{67.1} & \textbf{67.100} & \textbf{67.1}\\
			& PSO & 35.072 & 103.400 & 128.200 & 153.000 & 0.000 & 1.0 & 1.000 & 1.0 & 7.425 & 93.5 & 98.750 & 104.0\\
      	\bottomrule
    	\end{tabular}
    	\label{tab5:exp2_buoys}
	\end{sidewaystable}
	
	\begin{sidewaystable}
  		\centering
    	\caption{\textbf{C2K data set:} computational comparison of ACG, GAG, GRA and PSO at different support (sup) values.}
    	\begin{tabular}{c c | c c c c : c c c c : c c c c}
      	\multirow{2}{*}{\textbf{Sup}} & \multirow{2}{*}{\textbf{Alg}} &
      	\multicolumn{4}{c}{\textbf{Run-time (sec)}} &
      	\multicolumn{4}{c}{\textbf{No. of Gradual Patterns}} &
      	\multicolumn{4}{c}{\textbf{Memory (KiB)}}\\
		& & Std.Dev & Best & Mean & Worse & Std.Dev & Fewest & Mean & Most & Std.Dev & Min & Mean & Max\\
      	\toprule
      	0.5 & ACG & 0.987 & 29.250 & 30.297 & 31.210 & 8.737 & 12.0 & 21.667 & 29.0 & 0.577 & 199.0 & 199.667 & 200.0\\
			& GAG & \textbf{4.319} & \textbf{24.650} & \textbf{27.370} & \textbf{32.350} & 0.577 & 1.0 & 1.333 & 2.0 & 1.155 & 114.0 & 114.667 & 116.0\\
			& GRA & 12.156 & 389.700 & 400.167 & 413.500 & \textbf{0.000} & \textbf{604.0} & \textbf{604.000} & \textbf{604.0} & 0.000 & 208.0 & 208.000 & 208.0\\
			& PSO & 0.985 & 28.380 & 29.357 & 30.350 & 0.577 & 1.0 & 1.333 & 2.0 & \textbf{1.155} & \textbf{105.0} & \textbf{105.667} & \textbf{107.0}\\
		\hline
		0.6 & ACG & 2.823 & 18.700 & 21.637 & 24.330 & 2.309 & 10.0 & 11.333 & 14.0 & 0.000 & 199.0 & 199.000 & 199.0\\
			& GAG & \textbf{0.789} & \textbf{14.480} & \textbf{15.380} & \textbf{15.950} & 0.000 & 1.0 & 1.000 & 1.0 & \textbf{1.000} & \textbf{101.0} & \textbf{102.000} & \textbf{103.0}\\
			& GRA & 15.437 & 156.700 & 170.567 & 187.200 & \textbf{0.000} & \textbf{110.0} & \textbf{110.000} & \textbf{110.0} & 0.577 & 140.0 & 140.333 & 141.0\\
			& PSO & 5.184 & 13.930 & 19.403 & 24.240 & 0.000 & 1.0 & 1.000 & 1.0 & 12.038 & 95.1 & 102.100 & 116.0\\
		\hline
		0.7 & ACG & 1.194 & 20.870 & 22.210 & 23.160 & 1.528 & 2.0 & 3.333 & 5.0 & 1.732 & 198.0 & 199.000 & 201.0\\
			& GAG & \textbf{1.981} & \textbf{15.210} & \textbf{16.757} & \textbf{18.990} & 0.577 & 1.0 & 1.333 & 2.0 & 2.060 & 99.1 & 100.667 & 103.0\\
			& GRA & 15.473 & 126.000 & 143.867 & 152.800 & \textbf{0.000} & \textbf{28.0} & \textbf{28.000} & \textbf{28.0} & \textbf{0.000} & \textbf{80.8} & \textbf{80.800} & \textbf{80.8}\\
			& PSO & 4.457 & 20.180 & 22.793 & 27.940 & 0.000 & 1.0 & 1.000 & 1.0 & 9.597 & 95.9 & 106.967 & 113.0\\
		\hline
		0.8 & ACG & 150.867 & 23.680 & 178.360 & 325.100 & 1.732 & 0.0 & 2.000 & 3.0 & 1.732 & 199.0 & 201.000 & 202.0\\
			& GAG & \textbf{1.540} & \textbf{12.720} & \textbf{14.077} & \textbf{15.750} & 0.000 & 1.0 & 1.000 & 1.0 & 1.002 & 97.2 & 98.167 & 99.2\\
			& GRA & 2.799 & 94.620 & 96.863 & 100.000 & \textbf{0.000} & \textbf{12.0} & \textbf{12.000} & \textbf{12.0} & \textbf{0.000} & \textbf{65.3} & \textbf{65.300} & \textbf{65.3}\\
			& PSO & 3.041 & 19.580 & 21.730 & 23.880 & 0.000 & 1.0 & 1.000 & 1.0 & 13.789 & 92.5 & 102.250 & 112.0\\
		\hline
		0.9 & ACG & 0.566 & 15.510 & 16.157 & 16.560 & 0.000 & 0.0 & 0.000 & 0.0 & 0.000 & 199.0 & 199.000 & 199.0\\
			& GAG & \textbf{1.979} & \textbf{7.942} & \textbf{9.794} & \textbf{11.880} & \textbf{0.000} & \textbf{1.0} & \textbf{1.000} & \textbf{1.0} & 0.751 & 89.0 & 89.767 & 90.5\\
			& GRA & 3.938 & 35.560 & 38.913 & 43.250 & 0.000 & 0.0 & 0.000 & 0.0 & \textbf{0.058} & \textbf{40.7} & \textbf{40.767} & \textbf{40.8}\\
			& PSO & 0.842 & 11.370 & 12.313 & 12.990 & 0.000 & 1.0 & 1.000 & 1.0 & 3.404 & 91.7 & 94.700 & 98.4\\    	
      	\bottomrule
    	\end{tabular}
    	\label{tab5:exp2_c2k}
	\end{sidewaystable}
	
	
	
	\begin{sidewaystable}
  		\centering
    	\caption{\textbf{UCI data set:} comparison of ACG, GAG, GRA and PSO at different support (sup) values.}
    	\begin{tabular}{c c | c c c c : c c c c : c c c c}
      	\multirow{2}{*}{\textbf{Sup}} & \multirow{2}{*}{\textbf{Alg}} &
      	\multicolumn{4}{c}{\textbf{Run-time (sec)}} &
      	\multicolumn{4}{c}{\textbf{No. of Gradual Patterns}} &
      	\multicolumn{4}{c}{\textbf{Memory (KiB)}}\\
		& & Std.Dev & Best & Mean & Worse & Std.Dev & Fewest & Mean & Most & Std.Dev & Min & Mean & Max\\
      	\toprule
      	0.5 & ACG & 56.613 & 375.100 & 408.433 & 473.800 & \textbf{1.000} & \textbf{4.0} & \textbf{5.000} & \textbf{6.0} & 1.155 & 117.0 & 118.333 & 119.0\\
			& GAG & \textbf{99.090} & \textbf{297.500} & \textbf{367.400} & \textbf{480.800} & 0.577 & 1.0 & 1.333 & 2.0 & 1.550 & 96.5 & 97.633 & 99.4\\
			& GRA & 51.827 & 382.900 & 414.800 & 474.600 & 0.000 & 4.0 & 4.000 & 4.0 & \textbf{0.000} & \textbf{63.2} & \textbf{63.200} & \textbf{63.2}\\
			& PSO & 80.161 & 412.600 & 480.100 & 568.700 & 0.577 & 1.0 & 1.333 & 2.0 & 0.557 & 88.0 & 88.500 & 89.1\\
		\hline
		0.6 & ACG & 63.899 & 248.800 & 321.400 & 369.100 & \textbf{0.000} & \textbf{3.0} & \textbf{3.000} & \textbf{3.0} & 4.163 & 110.0 & 114.667 & 118.0\\
			& GAG & 15.357 & 263.800 & 280.700 & 293.800 & 0.000 & 1.0 & 1.000 & 1.0 & 0.503 & 95.4 & 95.867 & 96.4\\
			& GRA & 28.262 & 255.200 & 271.667 & 304.300 & 0.000 & 3.0 & 3.000 & 3.0 & \textbf{0.058} & \textbf{65.1} & \textbf{65.133} & \textbf{65.2}\\
			& PSO & \textbf{34.195} & \textbf{193.800} & \textbf{221.167} & \textbf{259.500} & 0.000 & 1.0 & 1.000 & 1.0 & 0.321 & 84.8 & 85.033 & 85.4\\
		\hline
		0.7 & ACG & 3.581 & 274.300 & 276.467 & 280.600 & 1.528 & 0.0 & 1.667 & 3.0 & 5.568 & 105.0 & 111.000 & 116.0\\
			& GAG & \textbf{16.633} & \textbf{240.300} & \textbf{251.833} & \textbf{270.900} & 0.000 & 1.0 & 1.000 & 1.0 & 0.361 & 95.2 & 95.500 & 95.9\\
			& GRA & 2.155 & 254.200 & 256.633 & 258.300 & \textbf{0.000} & \textbf{3.0} & \textbf{3.000} & \textbf{3.0} & \textbf{0.058} & \textbf{65.1} & \textbf{65.167} & \textbf{65.2}\\
			& PSO & 73.999 & 167.200 & 252.033 & 303.300 & 0.000 & 1.0 & 1.000 & 1.0 & 1.700 & 84.3 & 86.000 & 87.7\\
		\hline
		0.8 & ACG & 40.856 & 226.200 & 268.233 & 307.800 & 0.577 & 0.0 & 0.333 & 1.0 & 7.550 & 104.0 & 111.000 & 119.0\\
			& GAG & 53.166 & 224.400 & 266.433 & 326.200 & 0.000 & 1.0 & 1.000 & 1.0 & 0.321 & 95.7 & 95.933 & 96.3\\
			& GRA & \textbf{2.281} & \textbf{223.400} & \textbf{225.867} & \textbf{227.900} & \textbf{0.000} & \textbf{2.0} & \textbf{2.000} & \textbf{2.0} & \textbf{0.000} & \textbf{57.9} & \textbf{57.900} & \textbf{57.9}\\
			& PSO & 178.277 & 213.600 & 384.425 & 632.600 & 0.000 & 1.0 & 1.000 & 1.0 & 0.733 & 86.6 & 87.650 & 88.2\\
		\hline
		0.9 & ACG & 73.155 & 230.400 & 279.867 & 363.900 & 0.000 & 1.0 & 1.000 & 1.0 & 6.807 & 104.0 & 109.333 & 117.0\\
			& GAG & 9.498 & 227.700 & 234.200 & 245.100 & 0.000 & 1.0 & 1.000 & 1.0 & 0.153 & 95.7 & 95.833 & 96.0\\
			& GRA & \textbf{2.909} & \textbf{221.100} & \textbf{223.867} & \textbf{226.900} & \textbf{0.000} & \textbf{2.0} & \textbf{2.000} & \textbf{2.0} & \textbf{0.058} & \textbf{57.8} & \textbf{57.867} & \textbf{57.9}\\
			& PSO & 16.286 & 355.100 & 368.300 & 386.500 & 0.000 & 1.0 & 1.000 & 1.0 & 0.115 & 87.6 & 87.733 & 87.8\\    	
      	\bottomrule
    	\end{tabular}
    	\label{tab5:exp2_uci}
	\end{sidewaystable}
	
	\clearpage
	
	\subsubsection{Experiment 4: Consistent Gradual Patterns}
	\label{sec5.3.4:consistent_gps}
	This experiment identifies consistent GPs extracted by all algorithms from data sets: B\&C, C2K, Buoys with 15,000 tuples, UCI with 47,000 tuples. The results are shown in Table~\ref{tab5:consistent_gs}.
	
	\begin{table}[h!]
	\small
	\centering
	\caption{Consistent gradual patterns}
	\begin{tabular}{|l|l|}
		\hline
		Data set & Consistent gradual patterns\\
		\hline \hline
		B\&C & $\{(Insulin, \downarrow), (HOMA, \downarrow)\}$, support: 0.94\\
		\hline
		C2K & $\{(i1\_rcf\_1\_p, \uparrow), (i1\_rcf\_1\_e, \uparrow)\}$, support: 0.837\\
		\hline
		Buoys & $\{(Tz, \downarrow), (Tav, \downarrow)\}$, support: 0.945\\
		\hline
		UCI & $\{(Global\_activepower, \uparrow), (Global\_intensity, \uparrow)\}$,\\& support: 0.954 \\
		\hline
	\end{tabular}
	\label{tab5:consistent_gs}
	\end{table}

	\subsection{Discussion of Results}
	\label{sec5.4:discussion}
	
	\subsubsection{Memory Usage}
	\label{sec5.4.1:discuss_memory}
	We observe that generally ParaMiner and ACO-ParaM-iner algorithms have higher requirements for memory usage than ACO-GRAANK, GA-GP, GRAANK and PSO-GP algorithms. This phenomenon explains why ParaMiner and ACO-ParaMiner algorithms (1) yield \textit{`Memory Error'} outputs for relatively large data sets and (2) have slower run-times for relatively small data sets. As described in Section~\ref{sec2.3:gradual_traversal}, ParaMiner is based on a DFS strategy whose drawbacks (for the case of GP mining) are as follows:
	\begin{enumerate}
		\item a numeric data set has to be encoded into a transactional data set before DFS can be applied. This significantly increases the size of the data set and consequently size of usage memory required; and
		\item DFS employs a recursive approach to find all the longest paths. However, we notice from the experiment results that ACO-ParaMiner has improved run-time since it uses a non-recursive heuristic approach to achieve this.
	\end{enumerate}
	
	For the case of ACO-GRAANK, GA-GP, GRAANK and PSO-GP algorithms, their relatively low memory usages may be largely attributed to the modeling of tuple orders using a binary matrix. However, for relatively large data sets GRAANK algorithm also yields a \textit{`Memory Error'} output because it has to hold all possible candidate GPs' binary matrices in memory. ACO-GRAANK, GA-GP and PSO-GP algorithms have low memory requirements for relatively large data sets because they generate fewer GP candidates and learn to reject the useless candidates stochastically. This implies that they only store few binary matrices in memory. It should be noted that for data sets that can be handled by GRAANK algorithm, (for these data sets) ACO-GRAANK, GA-GP and PSO-GP algorithms have slightly higher memory usage requirements than the latter due to the stochastic process that they use to search for useful GP candidates.

	\subsubsection{Execution Run-time}
	\label{sec5.4.2:discuss_runtime}
	We observe that generally BFS-based algorithms (ACO-GRAANK, GA-GP, GRAANK and PSO-GP) have relatively faster run-times than DFS-based algorithms (\\ACO-ParaMiner and ParaMiner). This mainly is due to low memory usage requirements explained in Section~\ref{sec5.4.1:discuss_memory}. Also, we observe that ACO-GRAANK, GA-GP and PSO-GP algorithms have relatively faster run-times than GRAANK algorithm because they employ a heuristic approach that learns useful candidate GPs quicker than GRAANK's classical level-wise approach.
	
	For the case of GRAANK algorithm, we observe that in almost all the 4 data sets (B\&C, Buoys, C2K and UCI) the run-times reduce significantly as $\sigma$ increases from 0.5 to 0.9. This behavior may be attributed to the fact that the number of possible GP candidates always reduce as $\sigma$ increases. However, ACO-GRAANK, GA-GP and PSO-GP algorithms do not generate and test all possible GP candidates. Instead, they use stochastic processes to learn only the useful GP candidates and this phenomenon explains their relatively constant run-times as $\sigma$ increases.
	
	\subsubsection{Extracted Gradual Patterns}
	\label{sec5.4.3:discuss_gps}
	We observe that generally the number of GPs extracted by ParaMiner algorithm is relatively constant against different supports because it searches for all closed patterns (which correlates as many attributes as possible) \cite{Negrevergne2014}, while the number of GPs extracted by GRAANK algorithm varies significantly with support since it sear-ches for all frequent patterns (especially minimal and maximal patterns). ACO-GRAANK, GA-GP and PSO-GP algorithms ignore minimal patterns and heuristically search for frequent maximal GPs (which are fewer but each is composed of several minimal patterns) and ACO-ParaMiner stochastically searches for closed GPs. 
	
	The \textit{standard deviation} is generally greater for the case of ACO-GRAANK, ACO-ParaMiner, GA-GP and PSO-GP algorithms than for the case of GRAANK and ParaMiner algorithms. This may imply that GRAANK and ParaMiner algorithms are more stable since produce almost similar number of patterns for the same experiment at different runs.

	\section{Conclusions and Future Works}
	\label{sec6:conclusion}
	In this paper, we describe an ant colony optimization technique for BFS-based and DFS-based approaches for mining gradual patterns. Given the computational results in Section~\ref{sec5:experiments}, we establish that ACO-GRAANK and ACO-ParaMiner, GA-GP and PSO-GP algorithms out-perform their classical counterparts GRAANK and ParaMiner in terms of faster run-time and low memory usage requirements. However, this comes at the cost of extracting relatively fewer number of patterns.
	
	As can be seen in Section~\ref{sec5.3.3:computational}, the results reveal that generally \textit{standard deviation} of number of gradual patterns for ACO-GRAANK and ACO-ParaMiner is larger than that of their classical counterparts. This shows that these \textit{ant}-based algorithms are slightly unstable; therefore, they extract different number of gradual patterns at different runs. Future work may involve improving their stability.
		
	Another future work may involve extending the \textit{ant}-based approach to probabilistically select tuple portions of a data set in order to estimate dominant gradual patterns. At the present time all the tuples of a data set are used to determine frequent gradual patterns. Under those circumstances, the computational complexity of extracting gradual patterns is directly proportional to the number of tuples in the data set. In contrast, if portions of few tuples are used to estimate the frequent gradual patterns, the computational complexity may be reduced to a constant regardless of the the data set size.
	
	\section*{Acknowledgements}
	The authors would like to thank the French Government through the office of Co-operation and Cultural Service (Kenya) and the office of Campus France (Montpellier) for their involvement in creating the opportunity for this work to be produced. This work has been realized with the support of the High Performance Computing Platform: \textbf{MESO@LR}\footnote{\url{https://meso-lr.umontpellier.fr}}, financed by the Occitanie / Pyrénées-Méditerranée Region, Montpellier Mediterranean Metropole and Montpellier University.

	\section*{Conflict of interest}
	\begin{footnotesize}The authors declare that they have no conflict of interest.\end{footnotesize}
	
	\bibliography{main}

\end{document}